\documentclass{sig-alternate-05-2015}
\usepackage{graphicx}
\usepackage{amssymb,amsmath}

\usepackage{caption}
\usepackage{subcaption}
\usepackage{balance}
\usepackage{color}
\usepackage[hidelinks]{hyperref}
\PassOptionsToPackage{hyphens}{url}
\usepackage{breakurl}
\usepackage{xspace}
\usepackage{array}
\usepackage{booktabs}
\usepackage{mathtools}
\usepackage{pbox}
\usepackage{enumerate}
\usepackage{enumitem}
\usepackage{xcolor}
\usepackage{flushend}
\usepackage{pifont}
\usepackage{cite}
\usepackage{multirow}

\graphicspath{{./figures/}}
\newcommand{\client}{client}
\newcommand{\archname}{SALVE}
\newcommand{\latls}{\textsf{laTLS}}
\newcommand{\ladns}{\textsf{laDNS}}

\newcommand{\inlinetitle}[1]{\vspace{4pt}\noindent\textbf{#1}}

\definecolor{dygreen}{HTML}{2E7D32}

\newcolumntype{L}[1]{>{\raggedright\let\newline\\\arraybackslash\hspace{0pt}}m{#1}}
\newcolumntype{C}[1]{>{\centering\let\newline\\\arraybackslash}m{#1}}
\newcolumntype{R}[1]{>{\raggedleft\let\newline\\\arraybackslash\hspace{0pt}}m{#1}}

\makeatletter
\def\@copyrightspace{\relax}
\makeatother
\begin{document}

\CopyrightYear{2016} 
\setcopyright{acmlicensed}
\conferenceinfo{MobiCom'16,}{October 03 - 07, 2016, New York City, NY, USA}
\isbn{978-1-4503-4226-1/16/10}\acmPrice{\$15.00}
\doi{http://dx.doi.org/10.1145/2973750.2973766}

\title{\archname: Server Authentication with Location VErification}
\numberofauthors{5}
\author{
  \begin{tabular}{C{0.30\textwidth}C{0.30\textwidth}C{0.30\textwidth}}
  Der-Yeuan Yu & Aanjhan Ranganathan & Ramya Jayaram Masti \\[2pt]
  \affaddr{Department of Computer Science} & \affaddr{Department of Computer Science}  & \affaddr{Department of Computer Science} \\
  \affaddr{ETH Zurich} & \affaddr{ETH Zurich}  & \affaddr{ETH Zurich} \\
  \email{dyu@inf.ethz.ch} & \email{raanjhan@inf.ethz.ch}  & \email{rmasti@inf.ethz.ch}
  \end{tabular}
  \\[24pt]
  \begin{tabular}{C{0.35\textwidth}C{0.35\textwidth}}
  Claudio Soriente & Srdjan Capkun \\[2pt]
  \affaddr{Telef\'{o}nica Research} & \affaddr{Department of Computer Science} \\
  \email{claudio.soriente@telefonica.com} & \affaddr{ETH Zurich} \\
   & \email{capkuns@inf.ethz.ch}
  \end{tabular}
}

\maketitle

\begin{abstract}
The Location Service (LCS) proposed by the telecommunication industry is an architecture that allows the location of mobile devices to be accessed in various applications. We explore the use of LCS in location-enhanced server authentication, which traditionally relies on certificates. Given recent incidents involving certificate authorities, various techniques to strengthen server authentication were proposed. They focus on improving the certificate validation process, such as pinning, revocation, or multi-path probing. In this paper, we propose using the server's geographic location as a second factor of its authenticity. Our solution, SALVE, achieves location-based server authentication by using secure DNS resolution and by leveraging LCS for location measurements. We develop a TLS extension that enables the client to verify the server's location in addition to its certificate. Successful server authentication therefore requires a valid certificate and the server's presence at a legitimate geographic location, e.g., on the premises of a data center. SALVE prevents server impersonation by remote adversaries with mis-issued certificates or stolen private keys of the legitimate server. We develop a prototype implementation and our evaluation in real-world settings shows that it incurs minimal impact to the average server throughput. Our solution is backward compatible and can be integrated with existing approaches for improving server authentication in TLS.

\end{abstract}

%
%


%
%

%
%



\section{Introduction}

The Location Service architecture (LCS)~\cite{lcsarch} is a platform in the telecommunication infrastructure that disseminates location information of mobile devices.
It abstracts away specific methods used to estimate the location of devices within the network and provides a unified interface for third-party access.
Such information can be used in a variety of location-based services, such as social networking, navigation, localized advertising or emergency response. When estimated using secure localization methods, location further becomes a unique characteristic of the mobile device.

Location information obtained from mobile networks can be used in critical security applications. In particular,
we explore the adoption of location information provided by LCS as a second factor of web server authentication. This is motivated by the fact that security-critical online services (e.g., financial services~\cite{visaserver, mastercard}, e-health services~\cite{rodrigues2013analysis}) are often hosted at dedicated and physically-protected locations. The on-site protection makes it difficult or expensive for an adversary to infiltrate such data centers. This is especially true for remote attackers that are located far away from the target server. As a result, the server's presence in protected data center locations can be regarded as an additional factor of its authenticity.

The need for another factor of authentication stems from the limitations of current server authentication solutions. Currently, server authentication is achieved using Transport Layer Security (TLS) and public key certificates issued by Certificate Authorities (CAs).
However, recent incidents
of CA compromise~\cite{leyden2011diginotar,soghoian2012certified,
google2015maintaining, eckersley2011iranian} and flaws in the trust model
underlying the public key infrastructure~\cite{clark2013sok} have resulted in
server impersonation attacks~\cite{FakeSSLCert}.
To prevent such attacks, a wide range of solutions that focus on strengthening
server authentication have been proposed. These include approaches such as key
pinning~\cite{rfc7469, soghoian2012certified}, multi-path
probing~\cite{wendlandt2008perspectives, rfc6962},
certificate revocation~\cite{myers1998revocatoin, rfc6960}, short-lived
certificates~\cite{topalovic2012towards}, or channel-bound
credentials~\cite{channelid, karapanos2014effective}. These techniques mitigate
attacks where the adversary has a mis-issued certificate binding the server's
domain name to the adversary's key. However, they do not thwart an attacker that
learns the secret key of the server. Some of them, such as certificate
revocation or short-lived certificates, may at best reduce the time window of
attack.

We apply location-based authentication to strengthen the current certificate-based server authentication in TLS. In our approach, the server must prove its location to the
client in addition to presenting a valid certificate.
While the use of geographic location has been proposed for client
authentication~\cite{saroiu2009enabling,Lenders2008Location,luo2010veriplace},
it has not been applied to server authentication, which
poses a number of new challenges. In order to use location information for
server authentication, we identify the following three requirements:
\emph{(i)}~an infrastructure to securely estimate the current location of a
server, \emph{(ii)}~an infrastructure to securely collect and disseminate the
legitimate locations of a server, and \emph{(iii)}~a mechanism that allows a
\client{} to check that the current location of the server matches its legitimate location during a TLS handshake.

Our solution, \archname, is built on three existing technologies: the Location Service (LCS), Domain Name Service (DNS)~\cite{rfc1035} and TLS. We extend LCS to issue verifiable statements containing the location of a server. We use DNS to securely collect and disseminate the set of legitimate locations of a server, potentially deployed in different data centers globally. Using DNS to disseminate server location information is seamlessly integrated with the client since it already uses DNS to fetch the server's IP address. The use of DNSSEC further allows the client to verify the authenticity of DNS data.
Finally, we extend TLS to enable the client to verify the server's location statement, issued by LCS, and check that it matches the location fetched via DNS.
As a result, a remote attacker with a valid certificate but  who does not co-locate with the victim server is unable to impersonate it unless it further compromises LCS or DNS.
The design of \archname{} is backward compatible and requires only small modifications to existing solutions and the verification process is transparent to the user.

\inlinetitle{Contributions.}
In this paper, we make the following contributions. We present \archname{}, a novel
framework to authenticate a server using its geographic location as a second
authentication factor. Our solution resists server impersonation attacks where a
remote attacker obtains a fraudulent server certificate or the server's secret
key. We show that our approach can be realized using existing technologies such as DNS, LCS, and TLS.
We further demonstrate the feasibility of \archname{} by means of a prototype implementation and perform a large scale evaluation using PlanetLab~\cite{planetlab}. Our results show that \archname{} does not incur a significant impact to overall performance and only reduces server throughput (number of requests per second) by 4.3\% on average.
Finally, we discuss extensions of our approach and how it can be integrated with other existing solutions for improving server authentication in TLS.

\inlinetitle{Outline.}
The rest of this paper is organized as follows. We present our motivation and give a brief background on the Location Service architecture, the functions of DNS, and the TLS handshake in Section~\ref{sec:background}. In Section~\ref{sec:overview}, we describe the design of \archname{}, our location-based server authentication framework. We analyze the security of \archname{} in Section~\ref{sec:security-analysis}. In Section~\ref{sec:implementation}, we describe our prototype implementation. We further evaluate our implementation in Section~\ref{sec:evaluation}. In Section~\ref{sec:discussions}, we discuss possible extensions, optimizations and limitations of \archname{}. The conclusion is in Section~\ref{sec:conclusion}.

\section{Motivation and Background}
\label{sec:background}
We first motivate the potential and benefits of location-based server authentication.
We then review the main technologies used in this work: the Location Service architecture, DNS with DNSSEC, and TLS.

\subsection{Physically-Secured Data Centers}
There is a growing trend towards the use of dedicated and protected data centers for hosting web services, allowing server locations to be leveraged as a second factor for authentication. We analyze the top 20 websites listed by Alexa~\cite{alexa500} and find that they (e.g., Google~\cite{googleserver}) are all hosted in dedicated data centers. Furthermore, financial institutions like VISA~\cite{visaserver} and MasterCard~\cite{mastercard} use data centers with enhanced physical security. Standards like ISO/IEC 27001 for certifying the security measures of data centers have also been developed~\cite{iso27001}. In practice, data centers used for security-critical applications are additionally protected by several on-site security measures. For example, Google's data centers~\cite{googlewhitepaper} are protected using electronic access cards, perimeter fencing, biometrics, multifactor authentication, access logs, indoor and outdoor surveillance cameras, etc. Additionally, data centers are increasingly being built in remote areas to provide physical isolation and secure storage of sensitive data~\cite{swissdatacenter}.

In summary, as sensitive websites are attractive targets for the adversary, they are often deployed in private data centers with strong physical security measures.
This motivates the use of website location as a unique attribute of critical services that can be leveraged as an additional factor for server authentication.

\subsection{Location Service Architecture (LCS)}
\label{sec:lcs-arch}

\begin{figure}[t]
 \centering
 \includegraphics[width=\columnwidth]{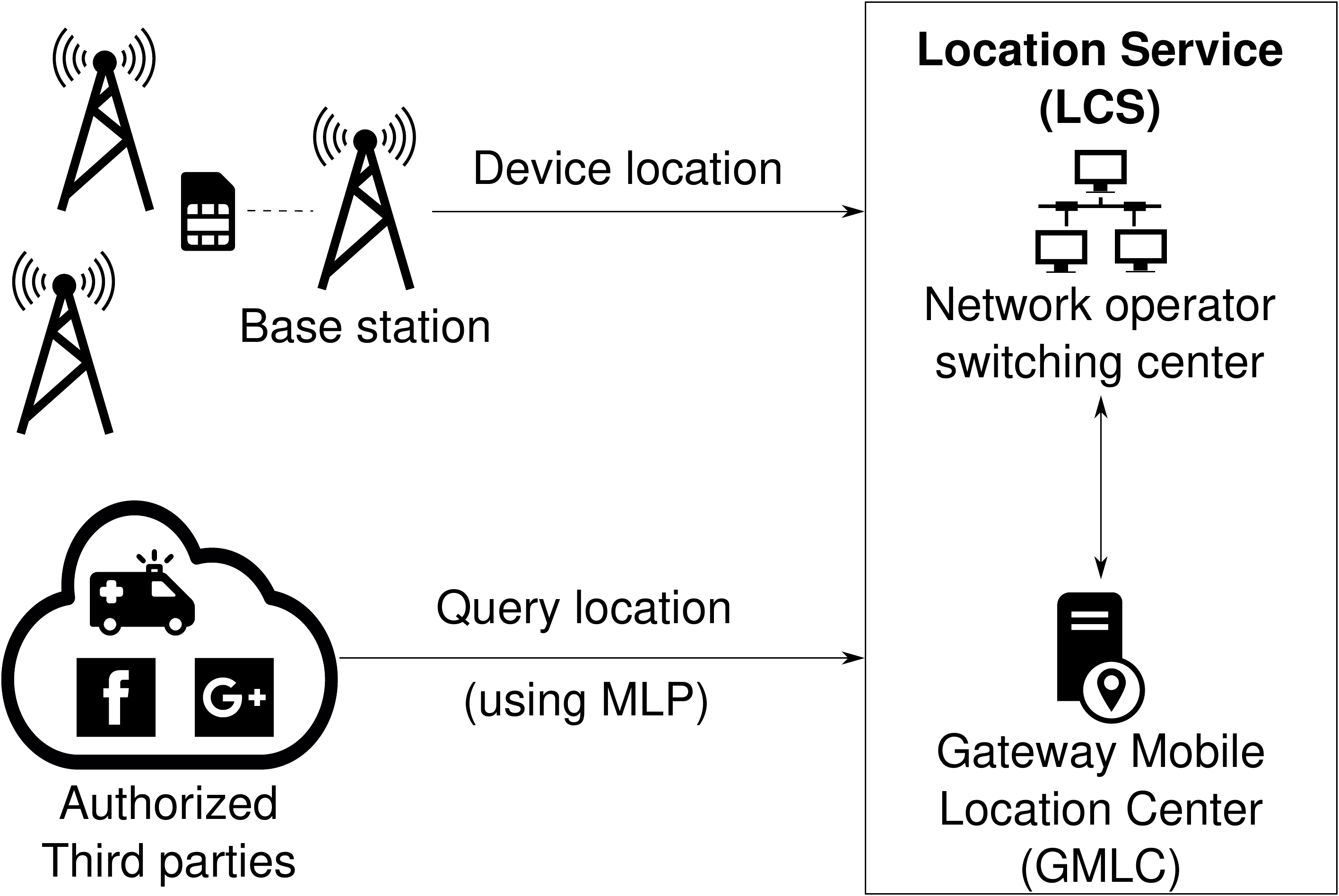}
 \caption{The LCS architecture. A SIM is localized by the telecommunication operator using its base stations. A third party queries the GMLC for the SIM's location using the Mobile Location Protocol. The GMLC enforces access control mechanisms to ensure that only specified parties are authorized to query for a specific SIM.}

 \label{fig:lcs}
\end{figure}

The Location Service architecture~\cite{lcsarch} allows the dissemination of location information about mobile devices. It enables a wide range of location-based services, such as emergency call positioning for rescue and support, location-based advertising, or tracking of assets and personnel. As shown in Figure~\ref{fig:lcs}, LCS consists of the following components: mobile devices with Subscriber Identity Modules (SIMs), base stations and network switching centers of telecommunication operators, the Gateway Mobile Location Center (GMLC), and various third parties or services. A mobile device has a SIM that base stations can localize (further detailed below). The operator stores the location information of each SIM at its network operator switching center. The location information is exposed through the GMLC, which is accessed by various location-based services or other third parties such as emergency services.

\inlinetitle{Localization Techniques.}
The telecommunication operator may use different techniques to localize a SIM
within its network, such as Cell IDs which identify the cell to which the SIM is connected~\cite{roxin2007survey}.
More accurate localization techniques such as U-TDOA~\cite{roxin2007survey} or
Enhanced Cell ID~\cite{ecellid} also exist. These techniques use multiple base
stations to measure the duration of signal transmission between the SIM and the
base stations to estimate its location. Alternatively, the base stations can measure received signal
power to localize the SIM. In this paper, we assume the presence of a localization system and that SIM locations are already available at the operator through the GMLC.

\inlinetitle{Mobile Location Protocol.}
Location-based services or other authorized third parties communicate with the GMLC using the Mobile Location Protocol~(MLP)~\cite{alliance2004mobile}.
MLP is an XML-based application layer protocol and is independent of the underlying physical network and the localization infrastructure. Using MLP, a user can query the GMLC for the location of a specific SIM. MLP supports query authentication, e.g., using passwords or pre-established session identifiers. The GMLC also implements access control to ensure that a user querying for a given SIM is authorized to do so. For example, the owner of a SIM may register and authorize himself with the GMLC or define which other parties are authorized to query for its own location.

\subsection{DNS and DNSSEC}
\label{sec:dnssec}
Browsers currently use DNS to obtain the IP addresses of websites to establish connections. IP address information is provided by website owners to DNS servers during the domain registration process. DNS is a hierarchical service that translates domain names to IP addresses in a recursive manner, referred to as \emph{DNS lookup}. A domain's IP address can be obtained from one or a group of DNS \emph{authoritative servers}. The client starts by querying a \emph{recursive server} for the IP address of a target domain, such as \texttt{www.example.com}. This recursive server then issues recursive queries to various DNS servers starting from the root authoritative server. This is followed by requests to a series of authoritative servers (in this example, the authoritative server of \texttt{.com} and then the one of \texttt{example.com}) until the IP address of the domain in question is obtained.

The responses of the DNS servers are organized as \emph{resource records}. Several types of resource records exist~\cite{rfc1035}. The most prominent among them is the \texttt{A} record that contains the IP address of the server for a given domain (\texttt{AAAA} for IPv6). If the \texttt{A} record is not available, the authoritative server provides a \texttt{NS} record which contains the IP address of the next authoritative server to be queried. Root DNS servers are maintained by ICANN, and other DNS servers are usually maintained by various Internet Service Providers or DNS registrars.


\inlinetitle{DNS Security Extensions.}
By design, DNS does not ensure the authenticity of resource records and is vulnerable to attacks such as DNS cache poisoning. To address this, DNSSEC~\cite{rfc4033, rfc4034, rfc4035} extends DNS with additional resource records used to authenticate existing records. DNSSEC assumes a public key infrastructure for the authoritative servers. Each authoritative server includes three additional types of records in its reply: \emph{(i)}~\texttt{RRSIG} records, containing the signatures on the other resource records, \emph{(ii)}~the \texttt{DNSKEY} record, containing the authoritative server's public key which can be used to verify the RRSIG records, and \emph{(iii)}~the \texttt{DS} record that includes the hash of the public key of a child domain's authoritative server (if any). The public keys of the authoritative servers form a chain of trust starting at the root server. For example, the root server supplies the \texttt{DS} record for the authoritative server of \texttt{.com} which, in turn, supplies the \texttt{DS} record for the authoritative server of \texttt{example.com}. The hash of the root server's public key must be known a priori by the querying client (e.g., browser) as a trust anchor.

\subsection{TLS Handshake}
\label{sec:background-tls}

A client connects to the server after learning its IP address from DNS.
To ensure communication security, the client and the server typically establish a secure session using TLS~\cite{rfc5246}. A TLS session is created through the process of a TLS handshake, which consists of the following steps: \emph{(i)}~the server sends its public key certificate to the client, who validates it to check the server's authenticity, \emph{(ii)}~the server optionally verifies the client's authenticity in the same way, and \emph{(iii)}~the two peers engage in a key exchange protocol to derive a shared \emph{master secret} for secure communication.\footnote{More specifically, the client and the server first agree on a \emph{pre-master secret}, which is then used to derive the master secret and subsequently discarded from memory.} The master secret is then used by both parties to generate symmetric session keys for encryption and authentication of all future communication within the TLS session.

\inlinetitle{Key Exchange Protocols.}
Various key exchange protocols exist for TLS. Since we later consider an
adversary who may steal the secret key of the server, we focus on the key
exchange variants of TLS that are based on ephemeral Diffie-Hellman key
exchange algorithms~\cite{diffie1992authentication} (detailed in
Section~\ref{sec:security-analysis}). These include TLS\_DHE
and TLS\_ECDHE, which are currently used by major online service providers like
Google~\cite{google-dh} and Twitter~\cite{twitter-pfs} because they provide
forward secrecy. The security of the implementation of Diffie-Hellman handshakes have been explored by Bhargavan et al.~\cite{bhargavan2014triple} and improved using the TLS Session Hash and Extended Master Secret Extension~\cite{ray2015transport}. In this work, using this extension, we assume that TLS master secrets are unique across different client-server sessions.

\section{\archname{}: Location-based Server\\Authentication}
\label{sec:overview}

\subsection{Overview}

We now introduce our approach for location-based server authentication.
We start by describing the system model, the set-up, and the attacker model.

\inlinetitle{System Model.}
We consider the typical scenario where a client (such as a browser) connects to
a web service (e.g., online banking) using TLS. The service can optionally be hosted at multiple servers deployed in different physically-protected data centers around the world.

\inlinetitle{Prerequisites.}
We assume that the legitimate IP address and location information of the web servers are provided by the domain owner to the DNS registrar. Legitimate locations can be stored in DNS using the existing \texttt{LOC} records~\cite{rfc1876}.
We further use DNSSEC to authenticate DNS records.
We also assume the existence of the LCS architecture that can be used to localize the web servers.
Each web server has a dedicated SIM\footnote{SIMs can be attached to servers through interfaces such as PCI-E or USB. Antennas can be installed outside the server room for better signal reception.} registered to LCS.
The telecommunication operator regularly localizes the SIMs and updates their locations in the switching center.
The GMLC is configured such that the legitimate web servers are the only parties authorized to query for the location of their own SIMs.
Communication with the GMLC occurs over a secure channel (e.g, using TLS).
The GMLC also has a public-private key pair, denoted by $sk_G$ and $pk_G$, respectively.
The client knows the GMLC's public key and the hash of the root DNS server public key. This can be achieved, for example, during the installation of the web browser or the operating system.

\begin{figure}[t]
  \centering
  \includegraphics[width=\columnwidth]{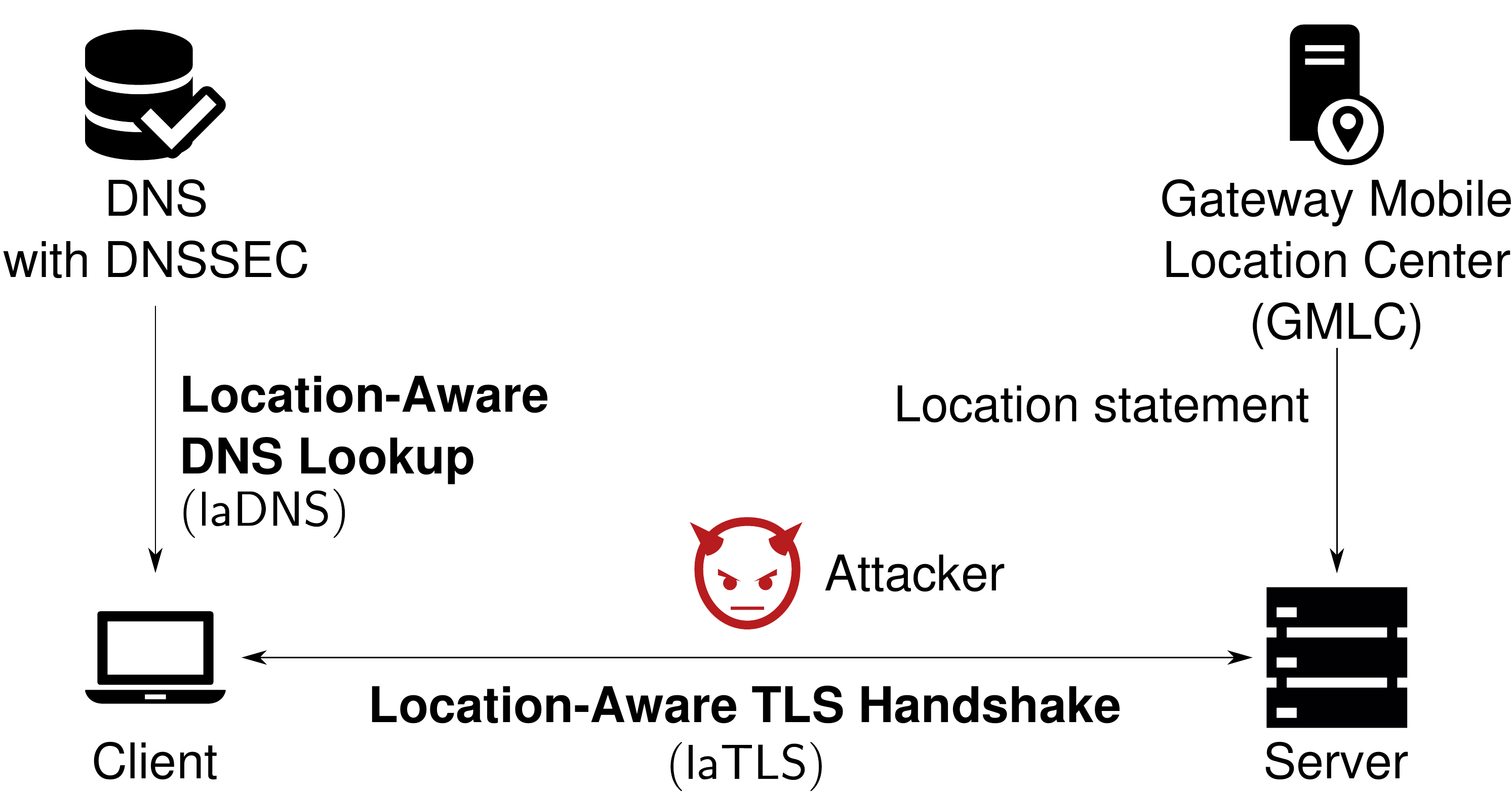}
  \caption{Overview of \archname{}. The client first
fetches the legitimate server location information from DNS using the
Location-Aware DNS Lookup. The client then uses this information to
authenticate the server during the Location-Aware TLS Handshake, which involves
the GMLC generating a location statement for the server. The attacker is remote and has access to the communication channel between the client and the server.}
  \label{fig:overview}
\end{figure}

\inlinetitle{Attacker Model.}
We consider a remote network attacker that has access to the channel between the client and the server.
The attacker has either compromised the secret key corresponding to the server's TLS certificate or obtained a valid TLS certificate for the server's domain through a compromised CA.

The attacker may also possess one or more SIMs registered with LCS.
The SIMs allow the adversary to ``appear'' at certain locations with respect to LCS.
Regarding his location, we assume the attacker to be \emph{remote} and does not co-locate with the server.
In other words, the location of the attacker's SIMs and the location of  any of the web server's SIMs are always different, according to LCS.
Finally, we assume that the attacker does not compromise LCS or DNS.

 \begin{figure*}[t]
  \centering
  \includegraphics[width=\textwidth]
  {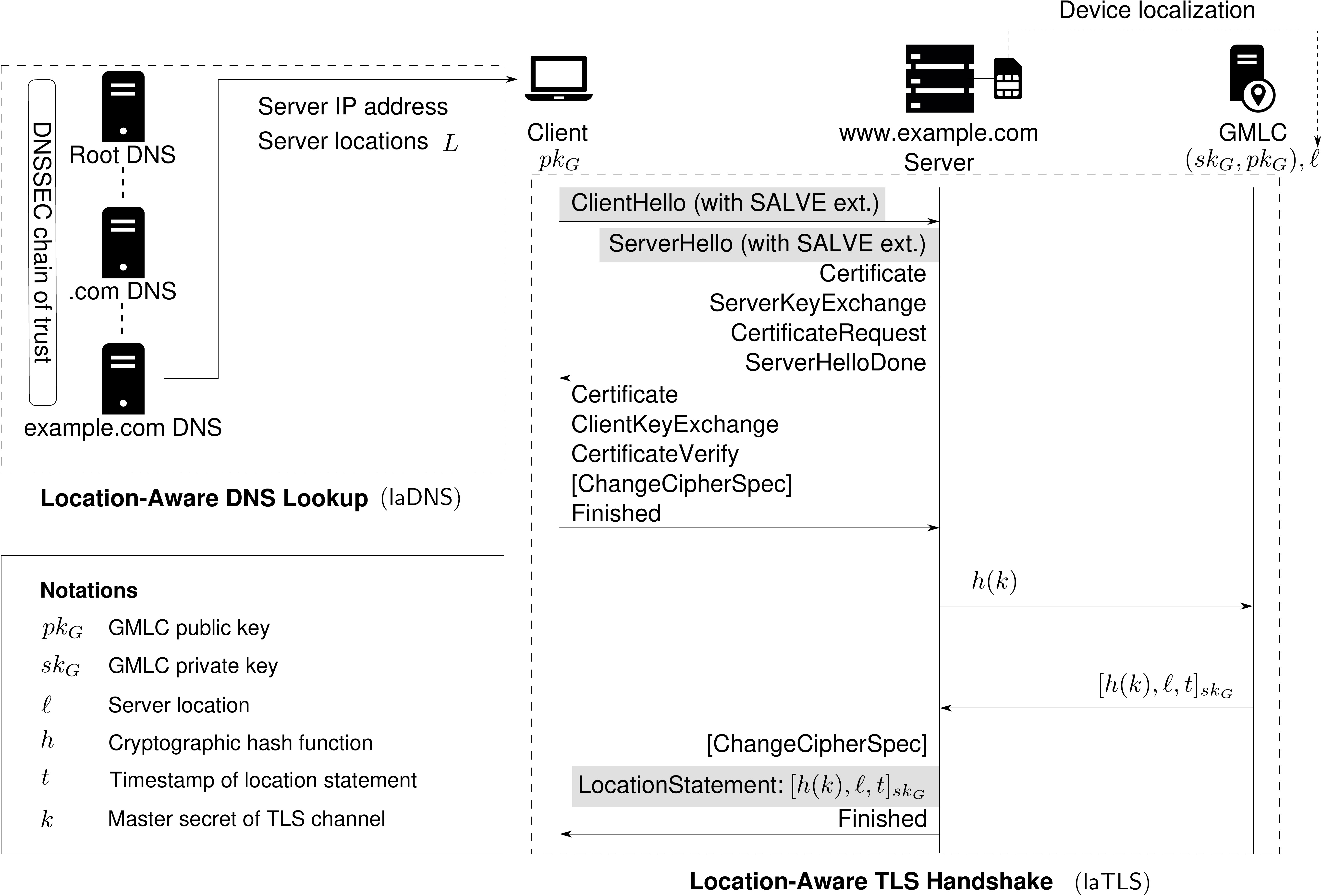}
  \caption{Location-based server authentication. \archname{} messages in \latls{} are shown in
gray. The Gateway Mobile Location Center (GMLC), as part of LCS, knows the
location of the SIM attached to the server. The client learns the legitimate
server locations through \ladns{}. After exchanging keys with the client, the
server sends a hash of the master secret $k$ to the GMLC to obtain a signed
location statement. This statement is sent as the \texttt{LocationStatement}
message to the client who verifies the server's location.}
  \label{fig:location-handshake-keyhash}
\end{figure*}

\inlinetitle{Solution Overview.}
As depicted in Figure~\ref{fig:overview}, \archname{} consists of two phases: \emph{Location-Aware DNS Lookup} (\ladns{}) and \emph{Location-Aware TLS Handshake} (\latls{}).

During the first phase, the client fetches the server's IP address and legitimate locations from DNS with DNSSEC enabled. This is integrated with the typical DNS lookup.

In the second phase, the client and the server engage in a modified TLS handshake, which involves the steps described below.
The client first receives and validates the server's certificate as in a standard TLS handshake. If the validation is successful, the client verifies the server's location as an additional authentication factor as follows.
During the handshake, the GMLC issues the server a signed \emph{location statement} that certifies the server's SIM location for this specific TLS session. The server forwards this statement to the client, who verifies its authenticity (using the public key of the GMLC) and matches the location contained in the statement against the server's legitimate locations fetched via \ladns{}.
We describe these phases in further detail below.

\subsection{Location-Aware DNS Lookup (\ladns{})}
During a Location-Aware DNS Lookup the client learns the legitimate server locations, in addition to the server's IP address. As shown in the left-hand side of Figure~\ref{fig:location-handshake-keyhash}, we extend the standard DNS lookup to also fetch all the legitimate locations of the servers of a given domain. Since the web service could be deployed at multiple data centers, the legitimate location information is composed of a set $L$ of possible server locations. Each location in $L$ is encoded in a \texttt{LOC} resource record. The \texttt{LOC} records are stored along with other records in the corresponding authoritative DNS server. We employ DNSSEC during the DNS lookup to ensure the authenticity of resource records. At the end of this phase, the client stores the legitimate server locations for later use.

We propose using DNS to store and disseminate legitimate server locations since it is already used to store their IP addresses. DNS is also used by current browsers when visiting websites and is deployed on a large scale.
The added attack surface is minimal since the number of DNS server keys relevant for a domain is small compared to the many CAs trusted by browsers~\cite{clark2013sok}.
Aside from DNS, \archname{} can also use other trustworthy online databases. Alternatively, browsers can store legitimate website locations in an offline manner, similar to certificate pinning~\cite{rfc7469}.

\subsection{Location-Aware TLS Handshake (\latls{})}
\label{sec:laTLS}

The Location-Aware TLS Handshake is a modified TLS handshake where the client verifies the server's location in addition to validating its certificate.
We design \latls{} as a TLS extension, which we later implement in modern TLS libraries (OpenSSL and NSS) as detailed in Section~\ref{sec:implementation}.
Messages exchanged by the parties are shown on the right-hand side of Figure~\ref{fig:location-handshake-keyhash} where gray boxes depict our additions to the standard TLS handshake.

The \texttt{ClientHello} message includes a custom extension type to indicate the request for \latls{}. Similarly, the \texttt{ServerHello} message contains the custom extension type to indicate the server's support. Other messages used to negotiate the master secret, denoted by $k$, remain unmodified. As mentioned in Section~\ref{sec:background-tls}, we consider the use of ephemeral Diffie-Hellman key exchange to provide forward secrecy.

Upon receiving the client's \texttt{Finished} message, the server requests the GMLC
to issue a location statement for its SIM using the MLP protocol. For efficiency, this request takes place over a long-term TLS
connection that lasts across different client connections. The request issued by the server includes the hash of
the master secret (i.e., $h(k)$) as an identifier of the client-server
handshake. The request also contains the server's credentials to access the
GMLC. The GMLC uses the credentials to check that the requester (i.e., the server) is authorized to
query for the SIM specified in the request, and aborts if this check fails. If
requester authentication succeeds, the GMLC replies with a signature, computed
using its private key $sk_G$, over $h(k)$ and the latest location information
of the server SIM. The location information in
Figure~\ref{fig:location-handshake-keyhash} is composed of the location denoted
by $\ell$ and  the time $t$ at which localization was performed. The server
receives the signed location statement and forwards it to the client in a new
TLS message type---\texttt{LocationStatement}.

Upon receiving the location statement, the client verifies the signature using
the public key $pk_G$ of the GMLC. The client also checks that the hash of the master
secret in the statement matches the one computed locally, and verifies that
the server is at a legitimate location according to the set $L$. The determination of whether the server location is legitimate is application-dependent. For example, the server's location in the location statement is legitimate if it matches (exactly or is within a given threshold distance from) one of the locations fetched from \ladns{}.
For example, the legitimate server location can be set to the center of the data center. The threshold can be set to less than a quarter of the dimensions of the premises to prevent an attacker at the edge from the data center from co-locating with the server.
The client also
checks for the timestamp $t$ in the location statement and rejects statements
where the time of the localization is older than a given threshold. If all these
checks are successful, the client accepts the server's authenticity and both peers start
exchanging application data. Otherwise, the client raises an alert
to indicate an error and quits the connection. We adopt the alert protocol in
TLS, used by clients to notify the server and quit the connection upon various errors, such as an invalid server certificate, a mismatch in the server's
\texttt{Finished} message, a decryption failure, etc.
To ensure the order of the messages, the client also raises an alert and terminates the connection if it receives \texttt{LocationStatement} message before the \texttt{Finished} message.

\section{Security Analysis}
\label{sec:security-analysis}

The adversary successfully impersonates a victim server if it obtains a location statement from the GMLC that the client accepts. Such a statement must include the hash of the master secret $h(k)$ expected by the client, a location that matches the set of legitimate server locations, and a timestamp that is not later than a pre-defined threshold.

In order to account for our defined adversary who has the secret key of the legitimate server, we focus on the DHE-variants of TLS (e.g., TLS\_DHE, TLS\_ECDHE). In the case of RSA-based key exchange (e.g., TLS\_RSA), the client chooses the master secret and sends it encrypted using the server's public key. This allows the adversary who knows the server's secret key to decrypt the message and obtain the master secret. The adversary then acts as a passive man-in-the-middle and lets the legitimate server carry out the entire handshake (including the location verification).
After the client authenticates the server, the adversary uses the master secret to derive the session keys of the TLS channel and effectively hijacks the connection.

DHE-based key exchange prevents a passive man-in-the-middle from learning the master secret agreed upon by the client and the server.
Therefore, the above attack would not work and, in the following, we only consider active man-in-the-middle attacks.

\subsection{Active Man-in-the-Middle}

An active man-in-the-middle attacker can engage in two TLS handshakes, one with the client (acting as the server) and one with the server (acting as the client).
This attacker therefore establishes two master secrets: one with the client (denoted by $k_{ca}$) and one with the server (denoted by $k_{as}$).

For simplicity, let $\ell_s$ be the legitimate server location (the one expected by the client)
and let $\ell_a$ be the adversary's location.
Since the adversary is remote, $\ell_s \neq \ell_a$.
During \latls{}, the client expects a location statement containing $[h(k_{ca}), \ell_s, t]$.
If the adversary  asks the GLMC for a location statement of its own SIM,
it receives $[h(k_{ca}), \ell_a, t]$, which would be rejected by the client because the location does not match the expected one.
If the adversary relays the location statement $[h(k_{as}), \ell_s, t]$ forwarded by the legitimate server,
the client rejects it because it does not contain the expected hash of the master secret.

Finally, we note that a location statement request carries the hash of the
master secret, so that the GMLC cannot learn the master secret and tamper with
the TLS channel between the client and the server.

\section{Implementation}
\label{sec:implementation}

We develop a proof-of-concept implementation of \archname{} by integrating and modifying existing software. Our modifications consist of roughly 1200 lines of code.
The implementation is composed of the following components: \emph{(i)}~a DNS server, \emph{(ii)}~the TLS libraries, \emph{(iii)}~the web server, \emph{(iv)}~the client, and \emph{(v)}~the GMLC.

\inlinetitle{DNS.} We use \texttt{bind9}~\cite{bind9} to set up a DNS server
to store an \texttt{A} and a \texttt{LOC} resource record for the web server.
We further enable DNSSEC support and populate all resource records with the
necessary DNSSEC resource records (\texttt{RRSIG} records) so that the DNS (particularly \texttt{LOC}) records can be authenticated. We use \texttt{delv} to issue \ladns{} queries from the client side for evaluation. No modifications are required here since the existing DNS and DNSSEC are used.

\inlinetitle{TLS Libraries.}
We implement \archname{} in the OpenSSL (version 1.0.2a) and the Network Security Services (NSS, version 3.18 RTM) libraries. For all TLS handshakes, we use the ephemeral elliptic curve Diffie-Hellman key exchange and use SHA256 as the cryptographic hash function for generating the master secret. For \latls{}, we use RSA with a 2048-bit key to sign and verify location statements. As a result, a location statement including its signature (encoded using Base64) is around 490 bytes in the format of MLP.

\inlinetitle{Web Server.}
We use the Apache HTTP Server 2.4.12 with the modified OpenSSL library that supports \latls{}. As mentioned in Section~\ref{sec:laTLS}, to reduce the overhead of re-establishing connections to the GMLC for every client connection, we modify Apache to maintain persistent TLS connections to the GMLC.

\inlinetitle{Client.}
We use two HTTP clients: a Chromium browser to evaluate client performance and ApacheBench (\texttt{ab}) for server benchmarks. We compiled Chromium (version 44.0.2388.0) with the modified OpenSSL library and ApacheBench (version 2.3) with the modified NSS library to support \latls{}.

\inlinetitle{GMLC.} We use RestComm GMLC~\cite{RestCommGMLC}, an existing GMLC implementation, and modify it to include our extension to issue signed location statements. The GMLC stores the location of the web server to issue location statements, which are represented in XML format based on the MLP standard.

\begin{figure*}[t]
  \centering
  \begin{subfigure}[b]{0.33\textwidth}
    \includegraphics[width=0.98\columnwidth]{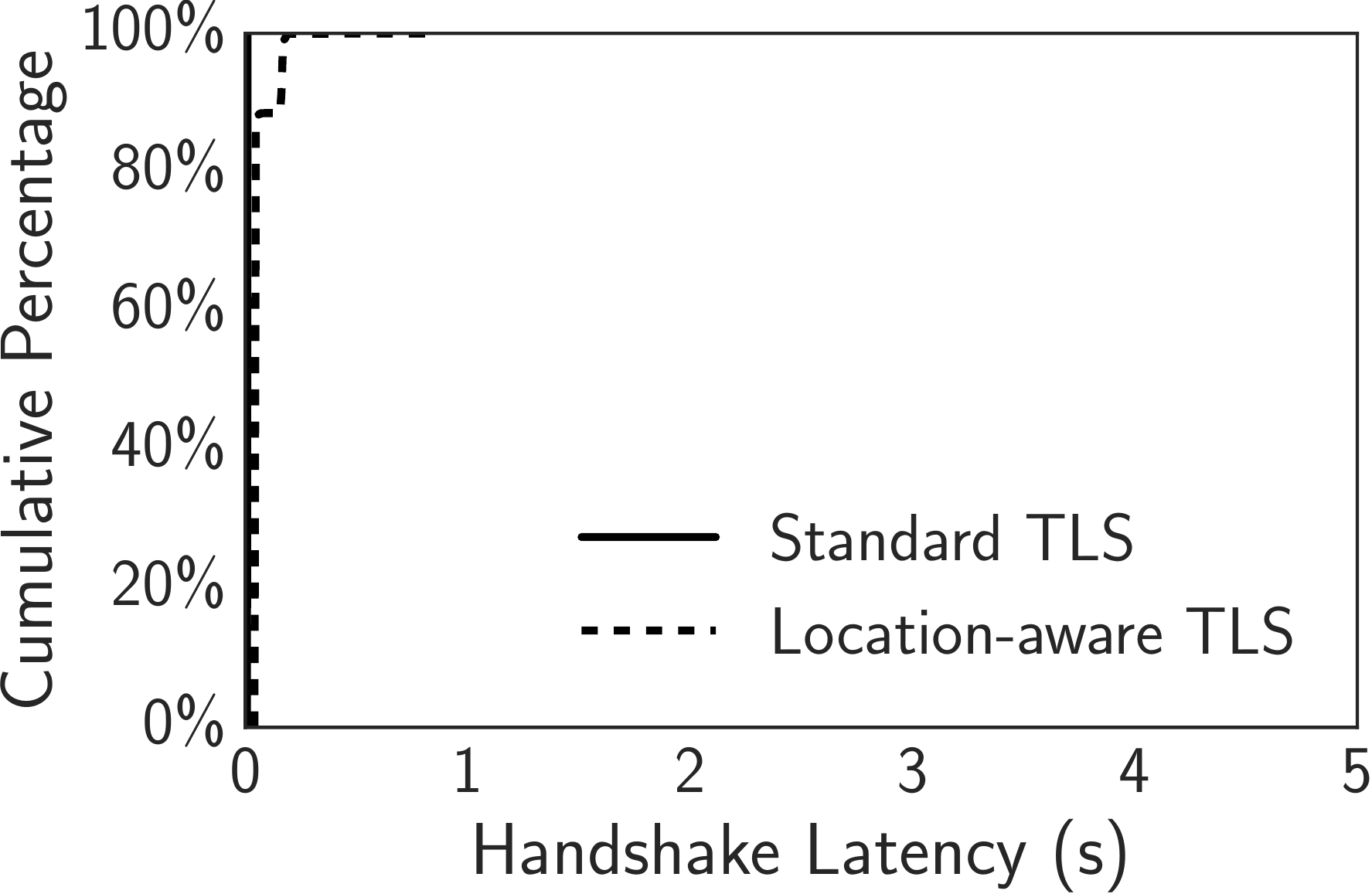}
    \caption{1 concurrent client}
  \end{subfigure}%
  \begin{subfigure}[b]{0.33\textwidth}
    \includegraphics[width=0.98\columnwidth]{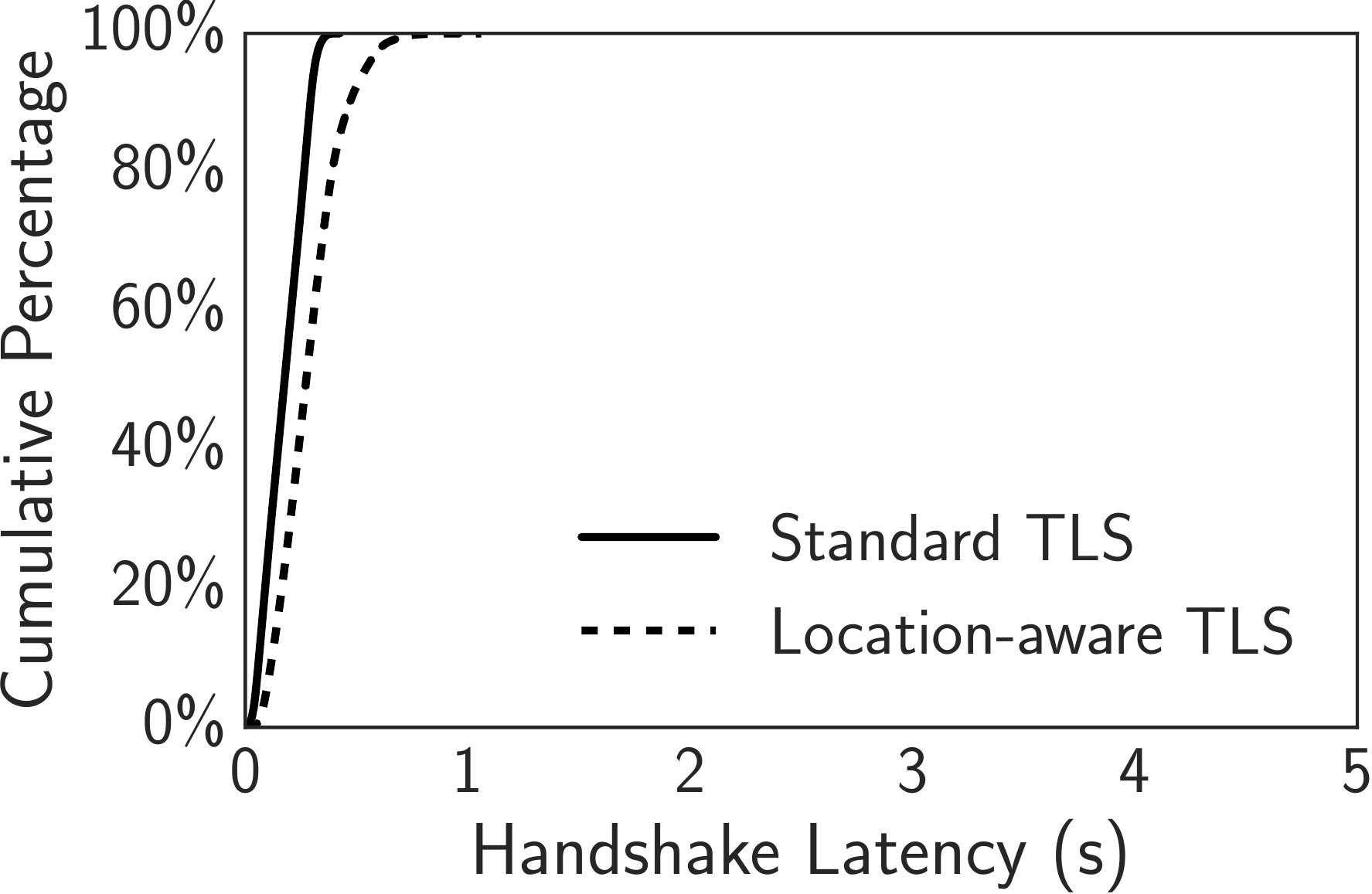}
    \caption{100 concurrent clients}
  \end{subfigure}%
  \begin{subfigure}[b]{0.33\textwidth}
    \includegraphics[width=0.98\columnwidth]{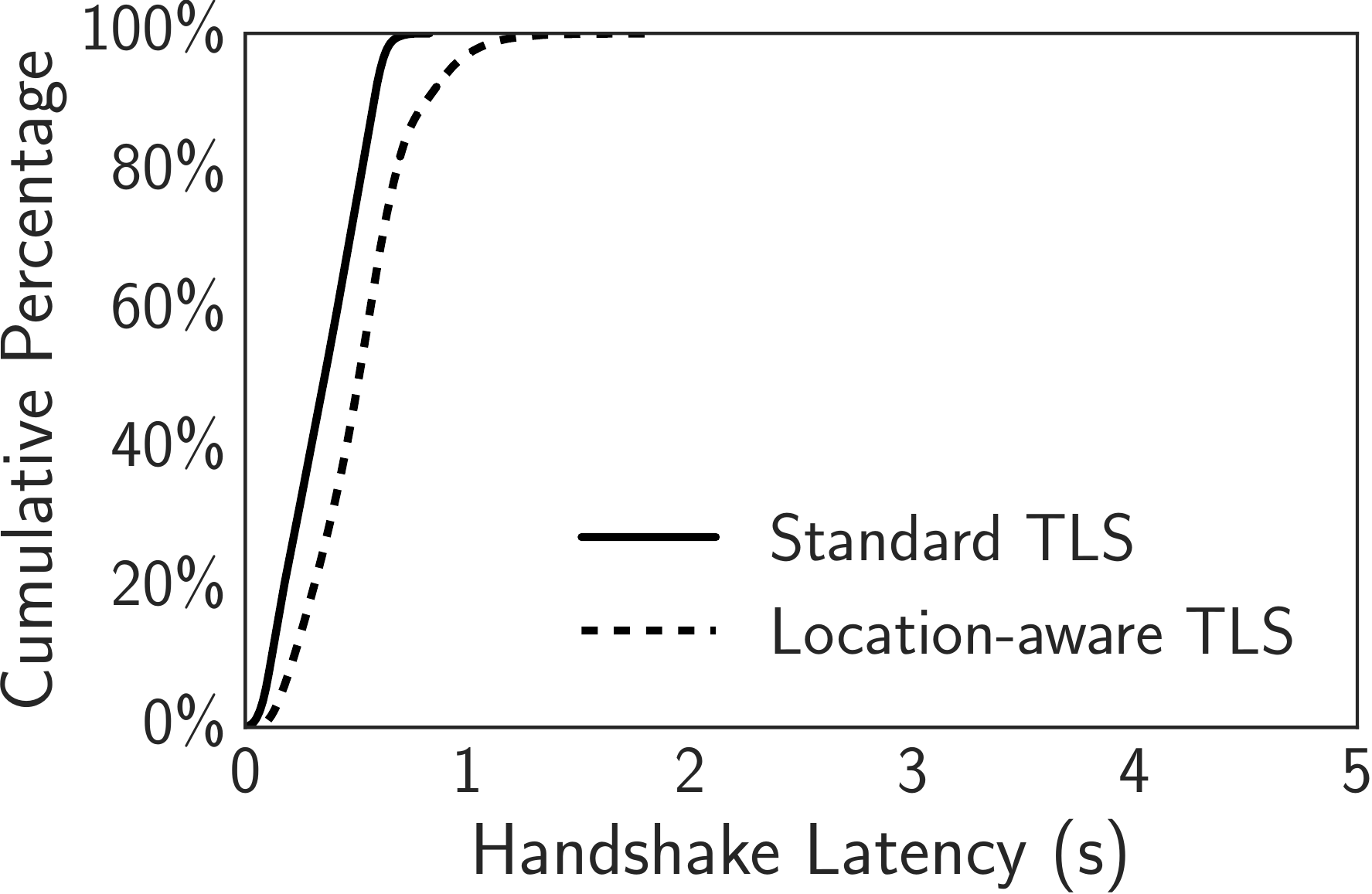}
    \caption{200 concurrent clients}
  \end{subfigure}%
  \\[6pt]
  \begin{subfigure}[b]{0.33\textwidth}
    \includegraphics[width=0.98\columnwidth]{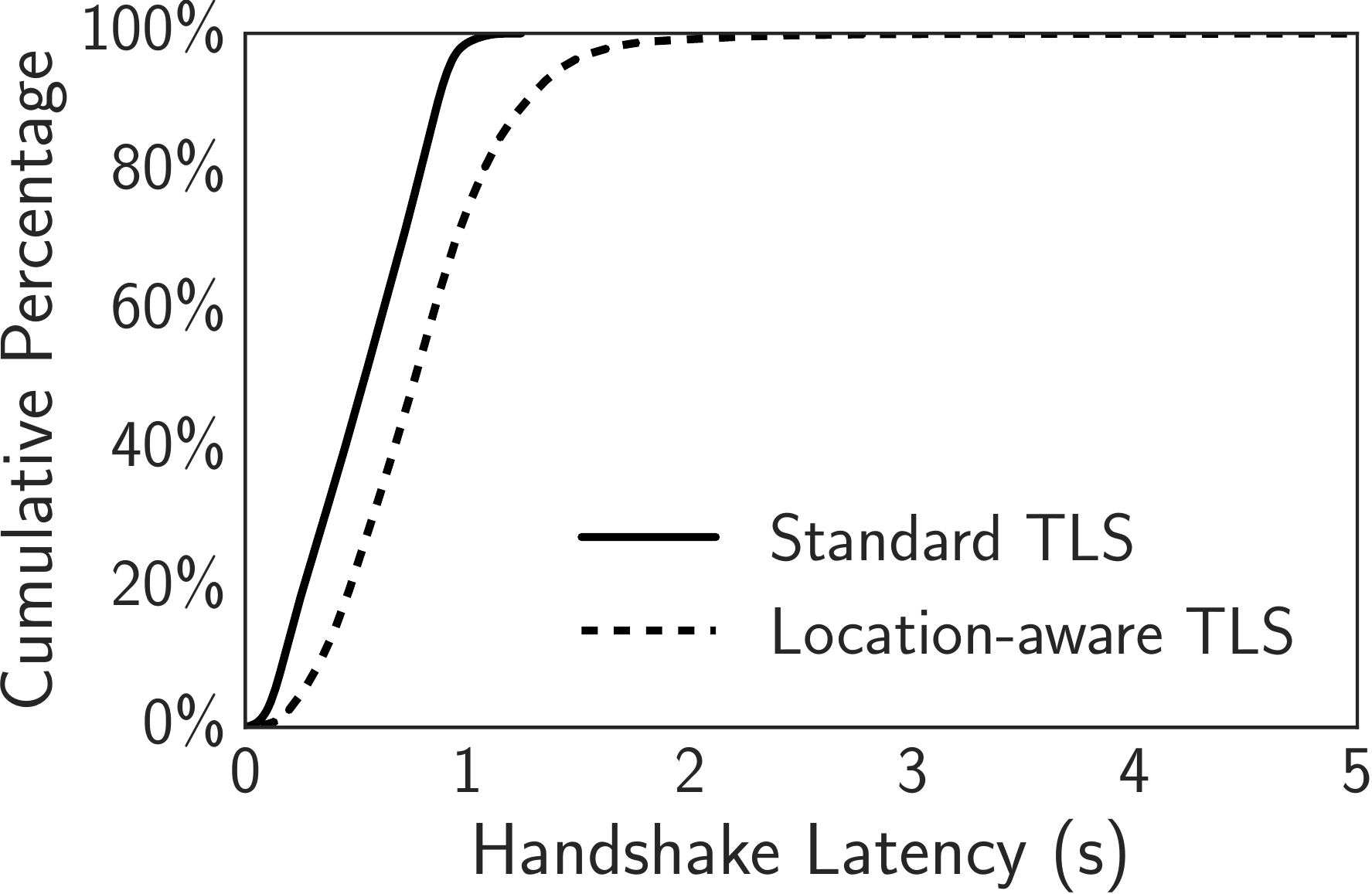}
    \caption{300 concurrent clients}
  \end{subfigure}%
  \begin{subfigure}[b]{0.33\textwidth}
    \includegraphics[width=0.98\columnwidth]{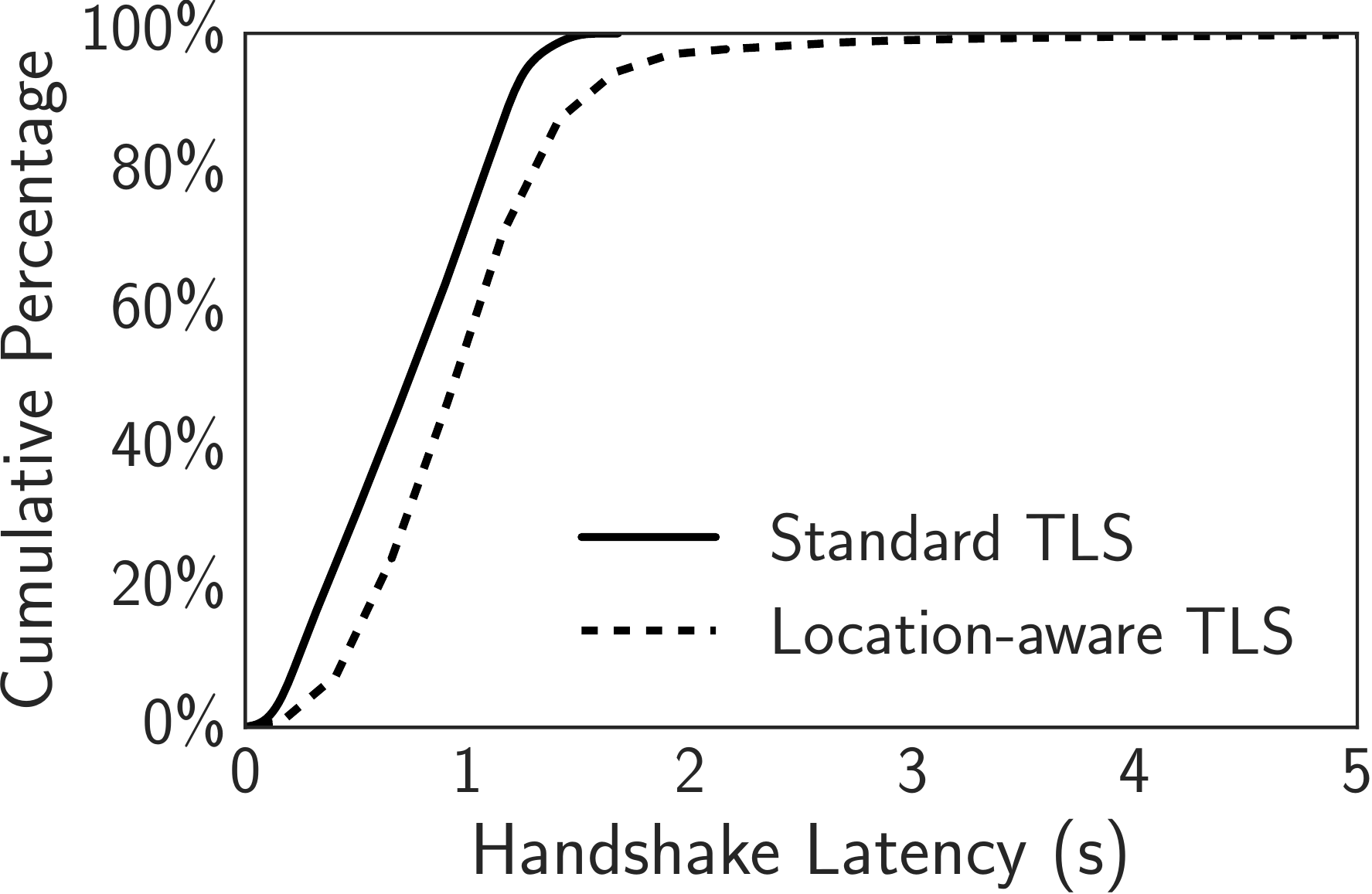}
    \caption{400 concurrent clients}
  \end{subfigure}%
  \begin{subfigure}[b]{0.33\textwidth}
    \includegraphics[width=0.98\columnwidth]{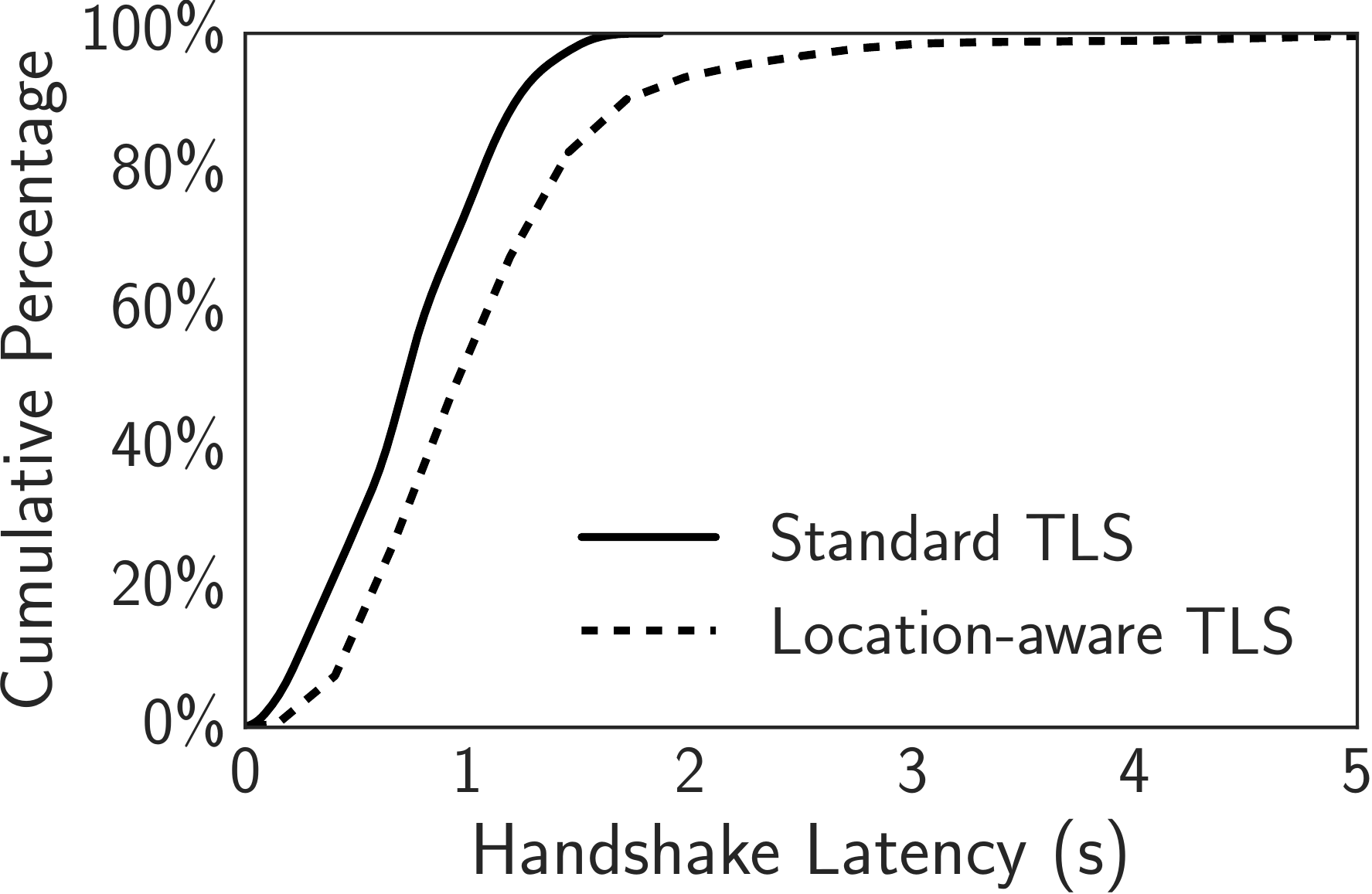}
    \caption{500 concurrent clients}
  \end{subfigure}%

  \caption{Cumulative distribution of server-side TLS handshake times for different numbers of concurrent clients evaluated locally on the server. \latls{} adds a delay to the distribution of the standard TLS handshake, which increases with the addition of concurrent clients (heavier load).}
  \label{fig:cdf-local}
\end{figure*}

\section{Evaluation}
\label{sec:evaluation}

We evaluate our implementation in a real-world deployment to demonstrate the feasibility of \archname{}.

\subsection{Goals and Methodology}

Our goal is to evaluate \archname{}'s overhead due to the modifications that the location-aware DNS lookup (\ladns{}) and the location-aware TLS handshake (\latls{}) entail.
We now introduce the deployment of \archname{} components, the experiments executed, and the metrics used to quantify the performance of the system.

\inlinetitle{Deployment.}
The DNS server runs on a machine with an Intel Core i7 CPU at 2.7~GHz and 4~GB RAM. We run the web server on a workstation with an Intel Core i7-4770 CPU at 3.40~GHz and 32~GB RAM in Switzerland. We envision that the GMLC, as part of the Location Service, is deployed on a large scale by telecommunication service providers, e.g., using a cloud platform. Therefore, we deploy a GMLC server on Amazon EC2 in Germany using the instance type \texttt{m4.xlarge}, consisting of a quad-core 2.4~GHz Intel Xeon E5-2676 processor and 16~GB RAM.

To reduce network effects, we first run the client locally on the same machine that hosts the web server. To observe the real-world performance impact on a large scale, we further deploy the client in Switzerland and in the following countries using PlanetLab: Germany (Goettingen), the US (Indiana) and Japan (Hiroshima).

\inlinetitle{Experiments.}
For \ladns{}, we use a locally-connected client to compare the performance of 200 lookups for the \emph{(i)}~the server's IP address and \emph{(ii)}~the server's IP address and location (encoded in a \texttt{LOC} record).

For \latls{}, we evaluate the server-side performance with clients running ApacheBench from different locations. The benchmark is performed with various levels of client concurrency (1 to 500 concurrent clients). Each benchmark configuration is executed on 50,000 fresh TLS handshakes or at least 10 minutes for both the standard TLS handshake and \latls{}. For the client-side performance, we run Chromium on the client to measure the overhead of verifying a location statement.
During evaluation, we compare the performance of \ladns{} and \latls{} against their respective vanilla solutions.

\inlinetitle{Performance Metrics.}
For our evaluation, we quantify the performance of \archname{} in terms of latency and throughput measured during its two phases: \ladns{} and \latls{}.
For \ladns{}, we focus on the latency experienced by the client when performing single lookups. More specifically, we measure the time between the client issuing a DNS request to the reception and validation of the retrieved records.

For \latls{}, we focus on three aspects: \emph{(i)}~the server-side latency by measuring the time between the server receiving a \texttt{ClientHello} to finally sending the \texttt{Finished} message at the end of the handshake (this includes the time for requesting the location statement from the GMLC and forwarding it to the client), \emph{(ii)}~the throughput by computing the average number of requests per second over a fixed period of time, and \emph{(iii)}~the client-side overhead for verifying the signature of the location statement and its contents.

\subsection{Location-Aware DNS Lookup}

We expect the \ladns{} to be efficient since it only requires the client to fetch the legitimate locations of the server while it fetches the IP address. Since our web server is deployed at a single location, the client fetches one \texttt{LOC} record during \ladns{}. The additional payload amounts to 420~bytes (30~bytes for the \texttt{LOC} record and 390~bytes for its corresponding \texttt{RRSIG} record, including record headers).

Our experiments show that the overhead of \ladns{} is reasonable compared to the time for a standard DNS lookup. A standard DNS lookup takes 36.7 milliseconds on average (standard deviation 4.81 milliseconds). A round of \ladns{} is 45.02 milliseconds on average (standard deviation 5.81 milliseconds), implying an increase of 8.3 milliseconds (23\%). Regardless of whether the DNS lookup includes location information, the DNS lookup time is typically amortized by caching the query results and using them in future connections, as long as the DNS records do not expire. DNS records expiration is defined in a time-to-live (TTL) field set by the authoritative DNS server. Given that typical DNS records can be valid for up to 5 days~\cite{rfc1912}, the overhead of \ladns{} is acceptable as it does not need to be performed for each client connection.

\subsection{Location-Aware TLS Handshake}

To reduce the effect of network latencies between the client and the server, we first analyze the results measured locally on the server machine. For a holistic view of server performance in a real-world setting, we further evaluate the performance with external client machines using PlanetLab.

\inlinetitle{Server-side Latency.}
We first show that \latls{} does not affect system stability in terms of the latency of TLS handshakes. To demonstrate this, we observe the cumulative distribution of handshake latencies measured by the local benchmark with different numbers of concurrent client connections. As shown in Figure~\ref{fig:cdf-local}, the distribution of \latls{} latencies is shifted to the right compared to the standard TLS handshake. The shift is due to the extra round-trip-time (RTT) spent by the server for requesting a location statement for each fresh client connection.

\begin{figure}[t]
  \centering
  \includegraphics[width=\columnwidth]
  {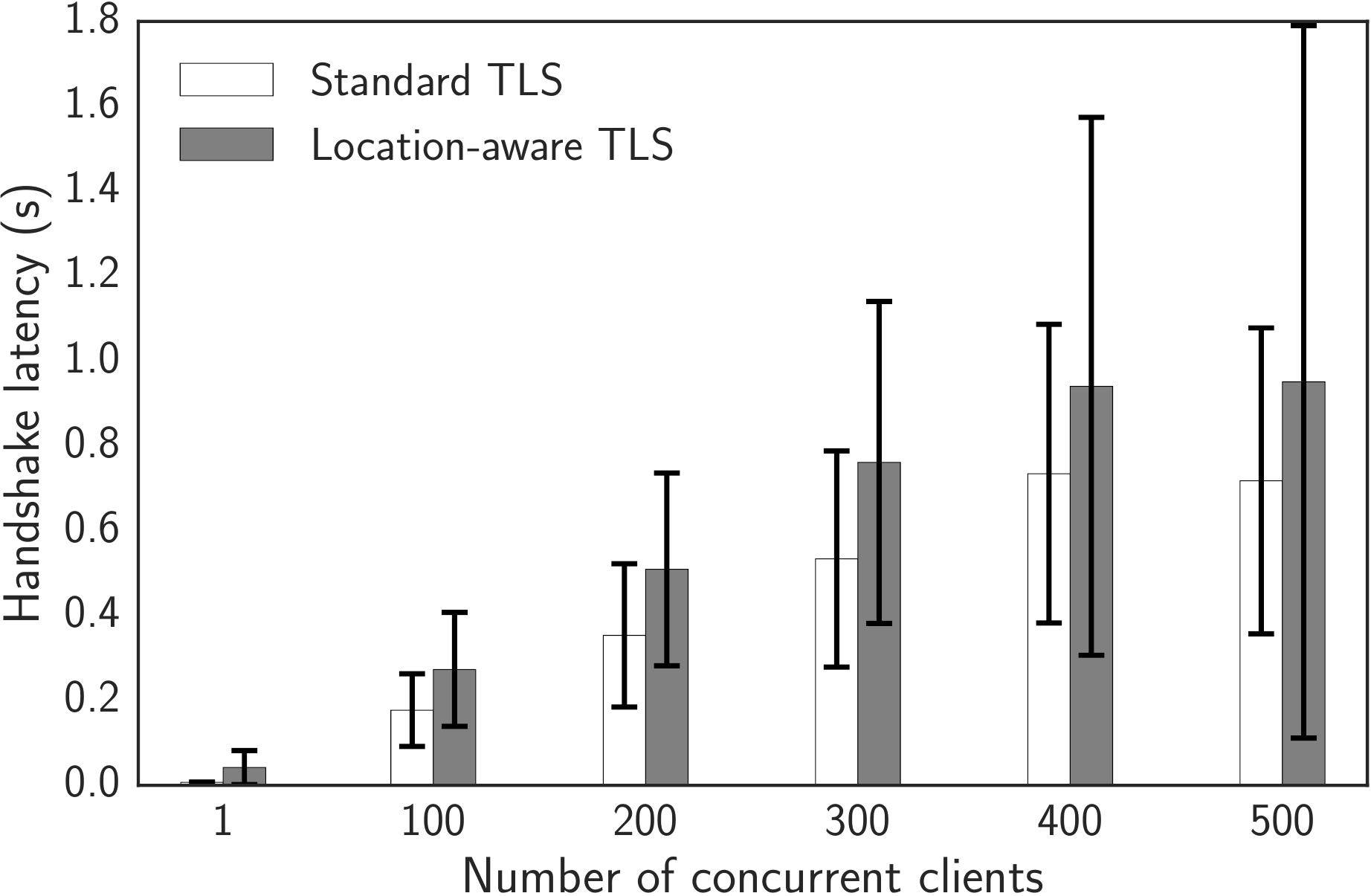}
  \caption{Server-side TLS handshake times for different numbers of concurrent clients evaluated locally on the server (median with standard deviation). The added latency (0.03 to 0.23 seconds) is primarily due to the round-trip communication between the web server and the GMLC.}
  \label{fig:latency-local}
\end{figure}

In further detail, \latls{} adds a minor overhead to the TLS handshake time. We compare its latency with the standard TLS handshake for various numbers of concurrent clients executed locally on the server (using ApacheBench), as shown in Figure~\ref{fig:latency-local}. \latls{} takes roughly 0.03 to 0.23 seconds more than a standard TLS handshake. As before, this is attributed to the round-trip communication between the server and the GMLC. Since web pages loading times are on the order of a few seconds~\cite{globalsitespeed}, such an overhead is not easily perceptible.



\begin{figure}[t]
  \centering
  \includegraphics[width=\columnwidth]
  {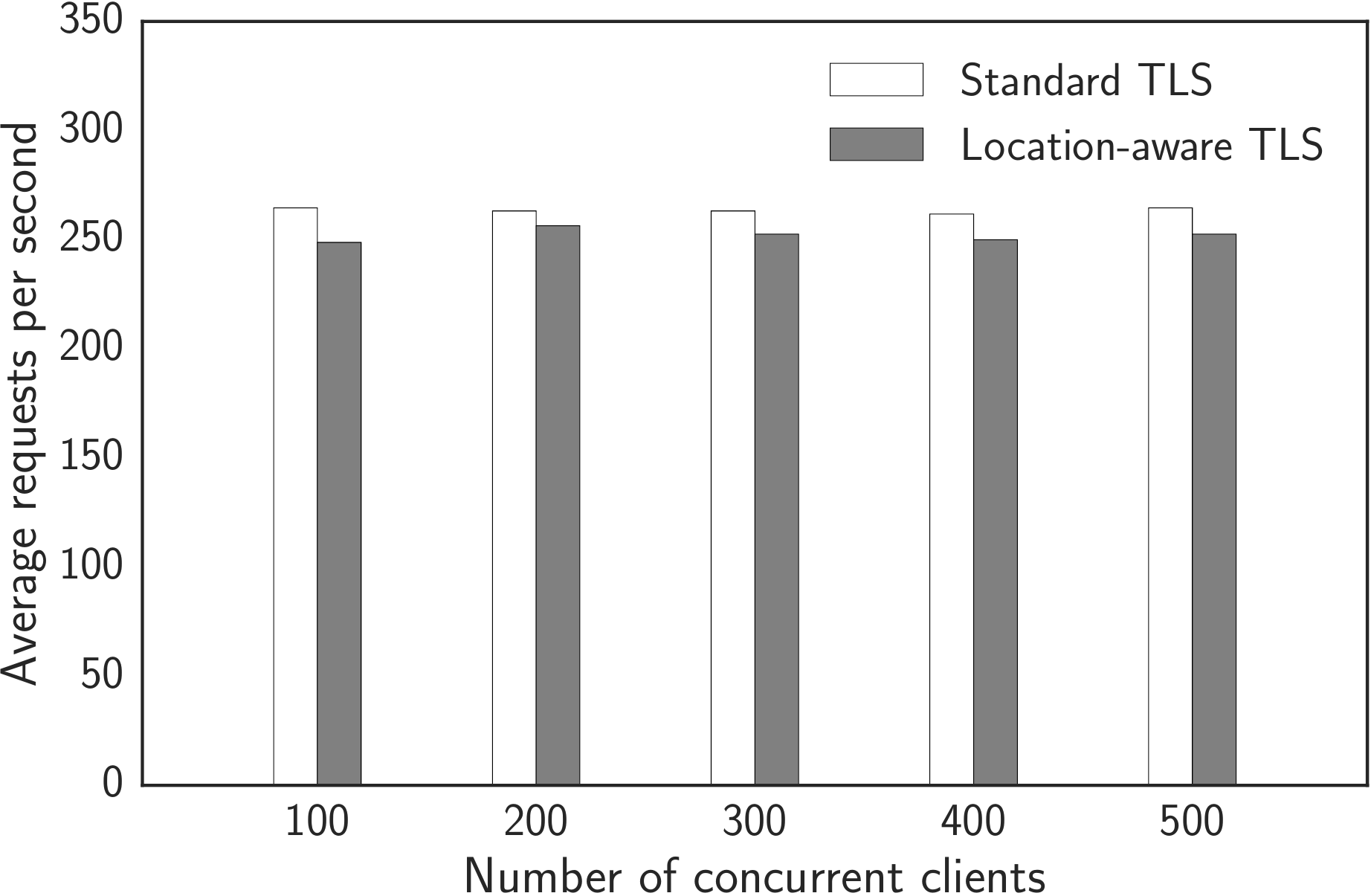}
  \caption{Server-side TLS handshake throughput for different numbers of concurrent clients evaluated locally on the server. The average number of requests per second is slightly reduced (2\% to 5\%) due to the round-trip communication between the web server and the GMLC.}
  \label{fig:throughput-local}
\end{figure}

\inlinetitle{Server-side Throughput.}
\latls{} does not significantly impact server throughput. We observe the number of requests the server handled during the benchmark time for each client concurrency level, as shown in Figure~\ref{fig:throughput-local}. The number of requests per second is reduced by an average of 4.3\% when compared to the standard TLS handshake.

\begin{figure*}[t]
  \centering
  \begin{subfigure}[b]{0.33\textwidth}
    \includegraphics[width=0.99\columnwidth]{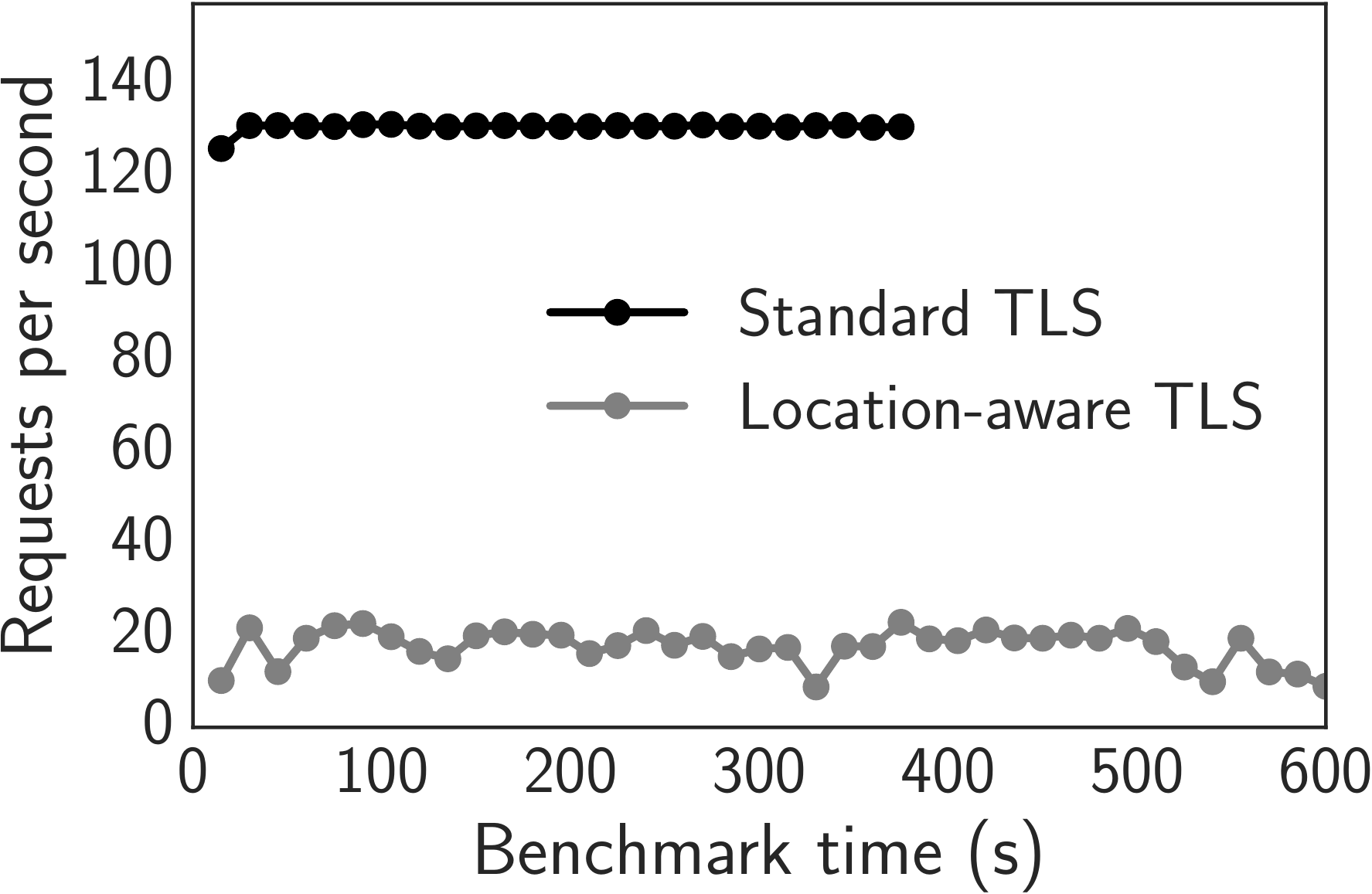}
    \caption{1 concurrent client}
  \end{subfigure}%
  \begin{subfigure}[b]{0.33\textwidth}
    \includegraphics[width=0.99\columnwidth]{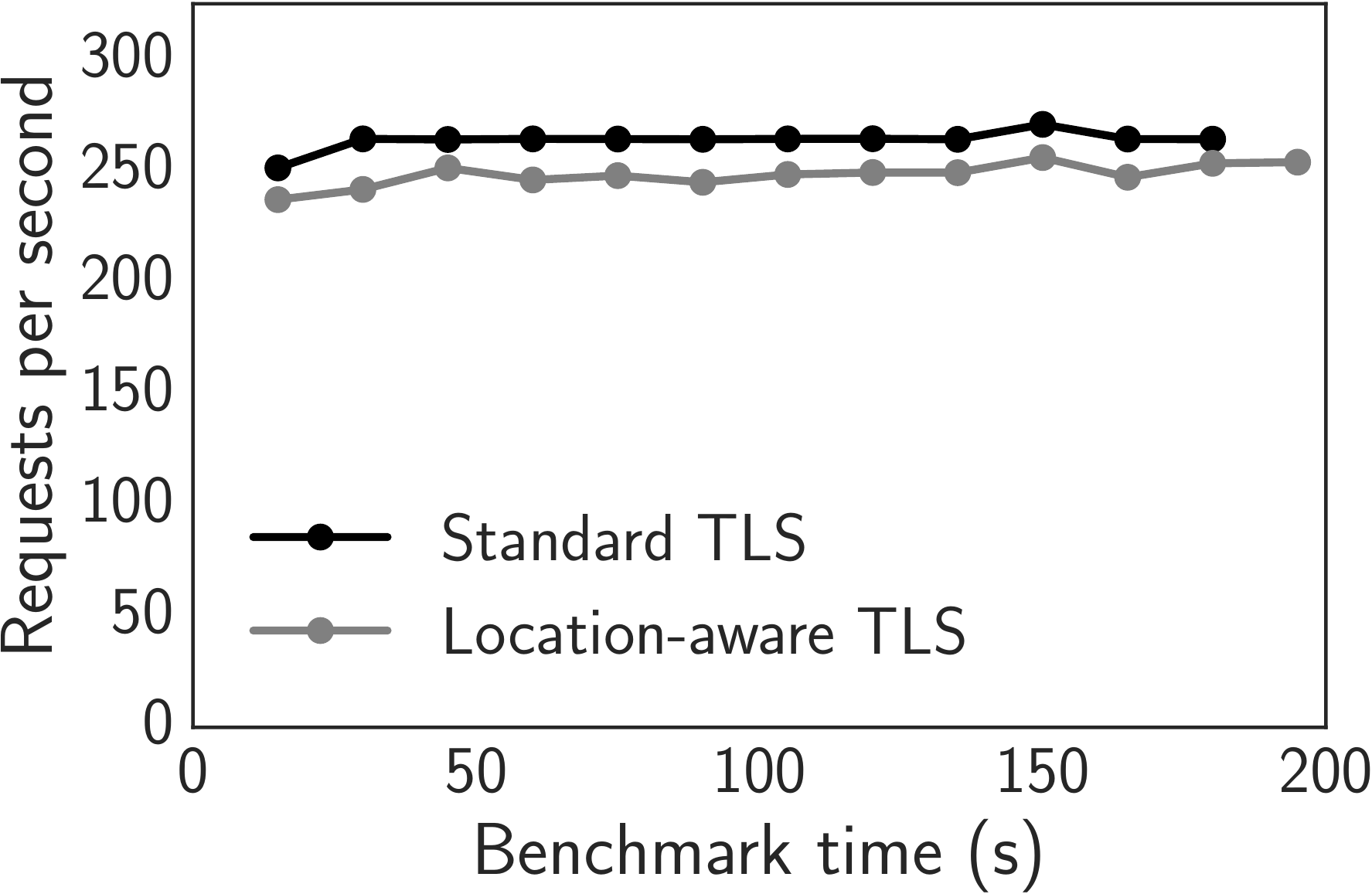}
    \caption{100 concurrent clients}
  \end{subfigure}%
  \begin{subfigure}[b]{0.33\textwidth}
    \includegraphics[width=0.99\columnwidth]{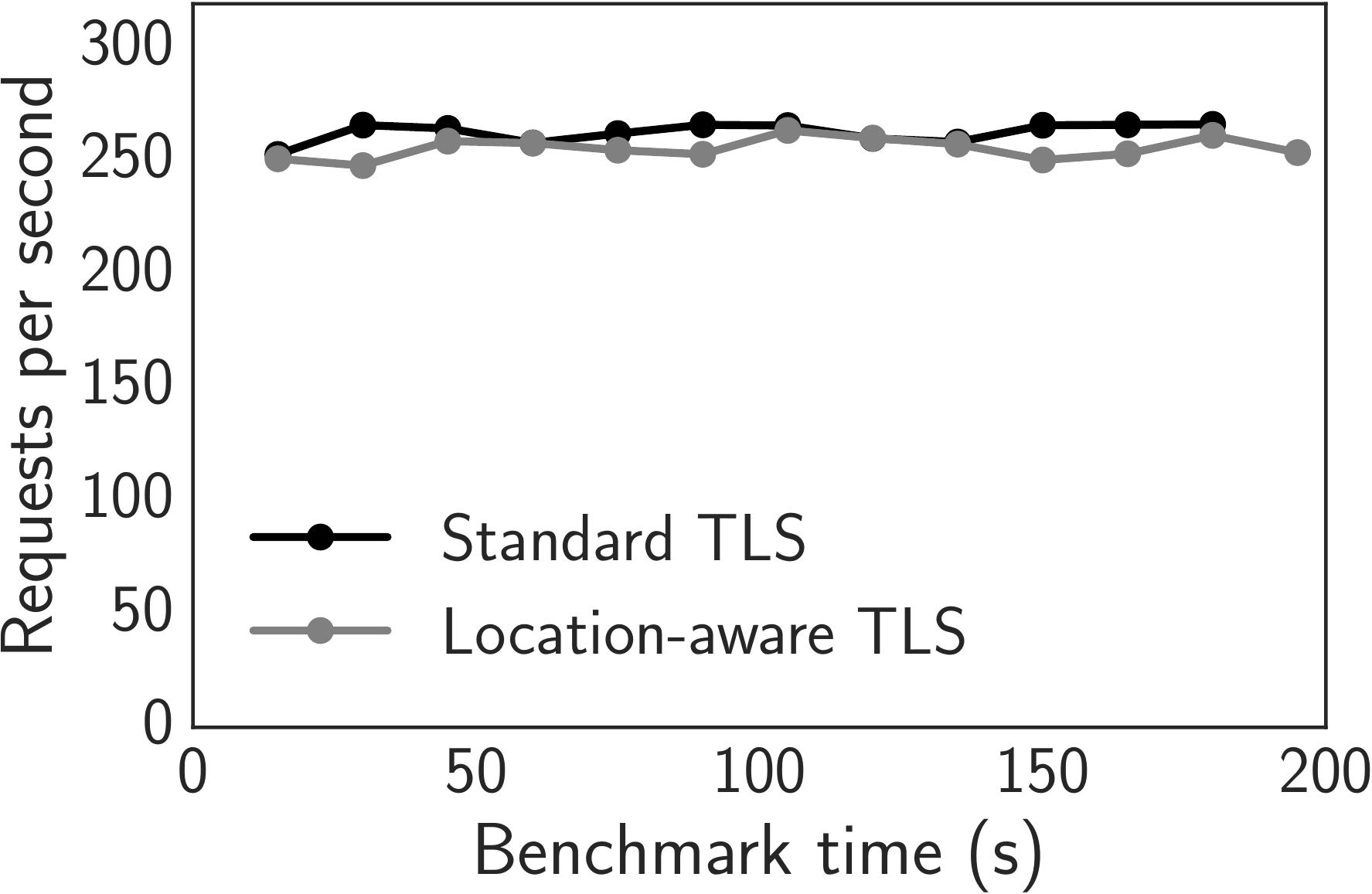}
    \caption{200 concurrent clients}
  \end{subfigure}%
  \\[6pt]
  \begin{subfigure}[b]{0.33\textwidth}
    \includegraphics[width=0.99\columnwidth]{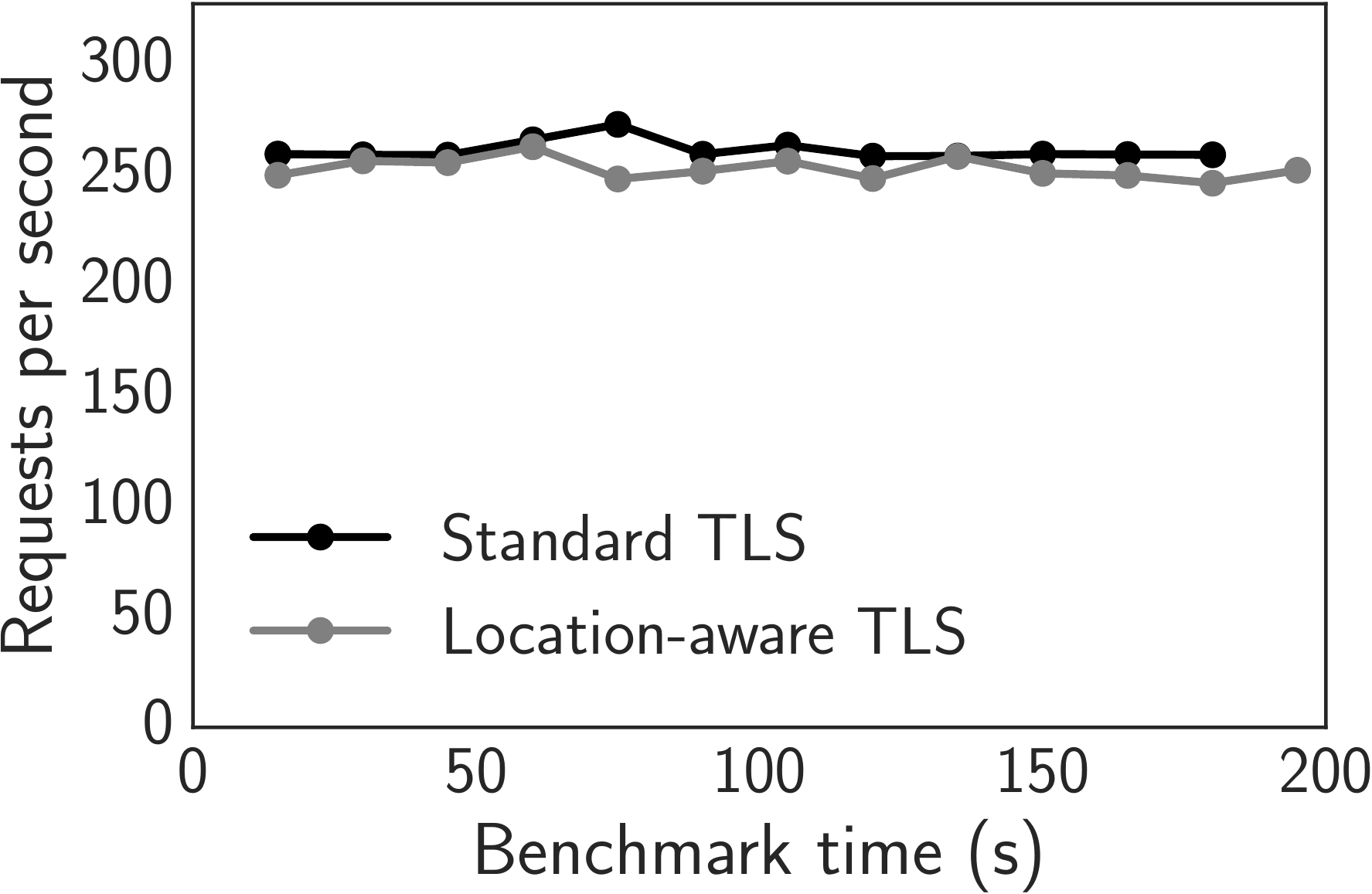}
    \caption{300 concurrent clients}
  \end{subfigure}%
  \begin{subfigure}[b]{0.33\textwidth}
    \includegraphics[width=0.99\columnwidth]{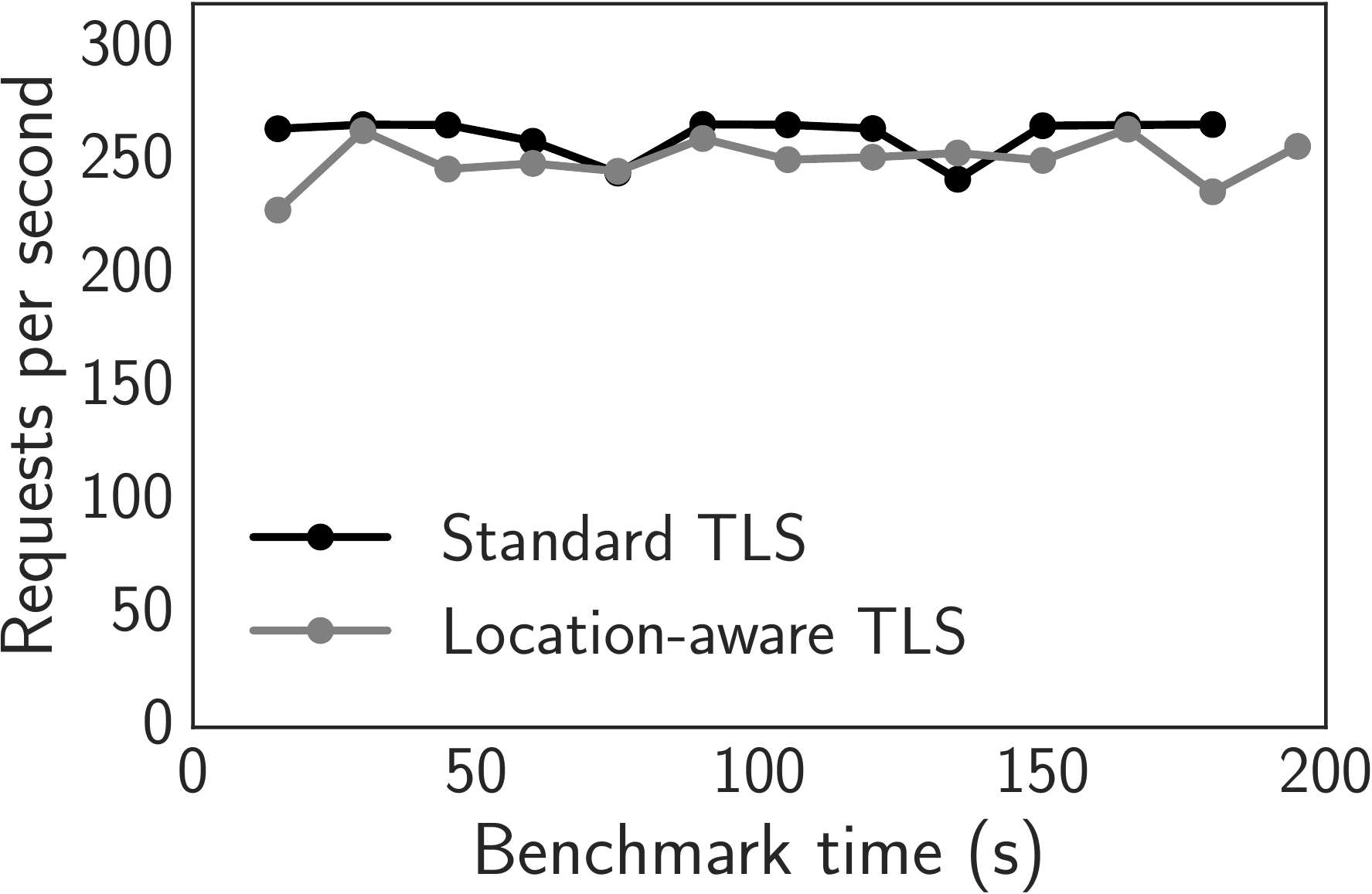}
    \caption{400 concurrent clients}
  \end{subfigure}%
  \begin{subfigure}[b]{0.33\textwidth}
    \includegraphics[width=0.99\columnwidth]{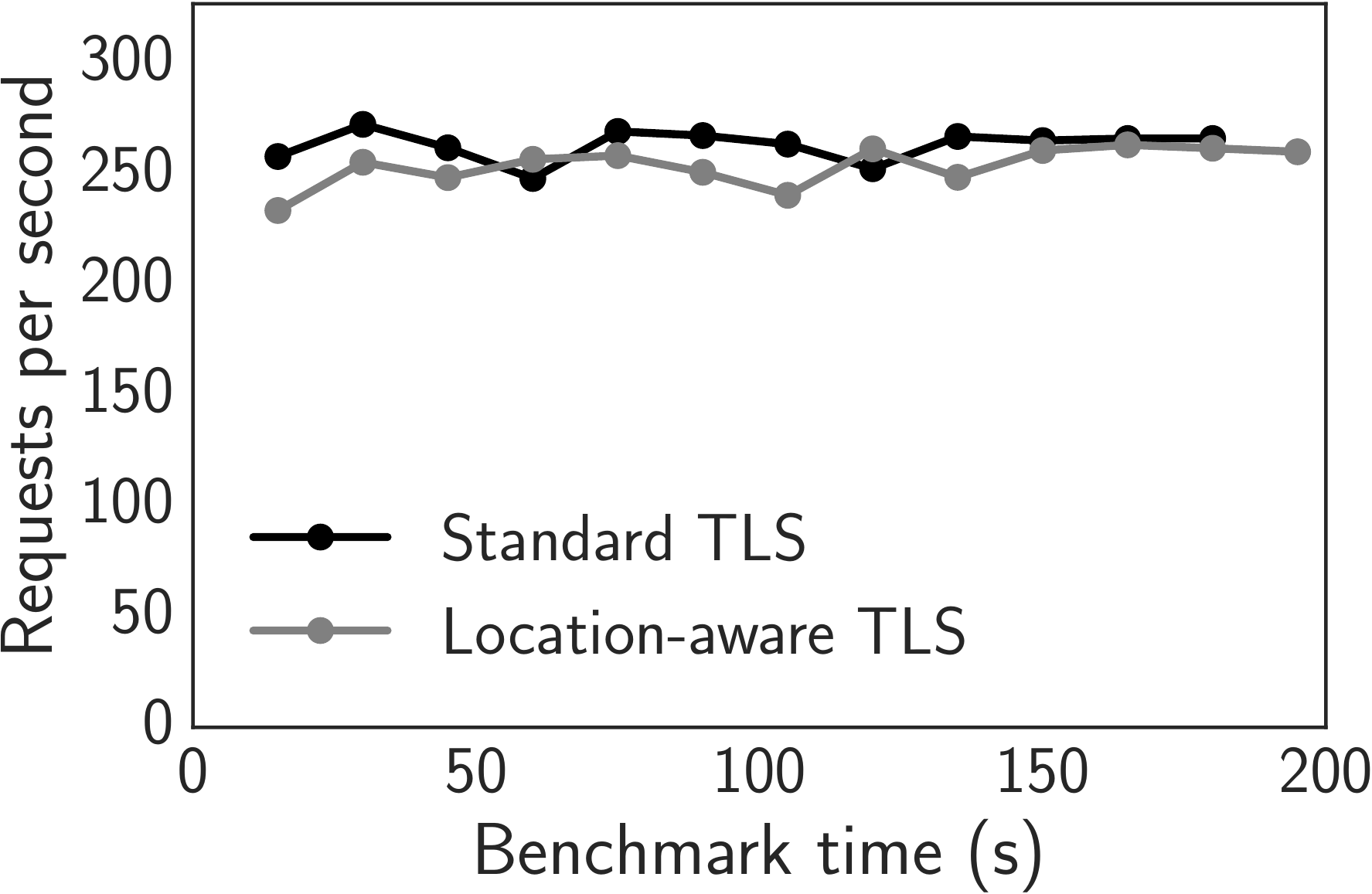}
    \caption{500 concurrent clients}
  \end{subfigure}%

  \caption{Server-side TLS handshake throughput for different number of concurrent clients evaluated locally on the server over 180 seconds, during which we observe stable performance. For all client concurrency levels, \latls{} incurs a minor reduction on the number of client requests that the server can handle concurrently.}
  \label{fig:throughput-local-stability}
\end{figure*}

The server-side throughput is also stable across time. We measure the average requests per second every 15 seconds and observe a stable reduction between \latls{} and the standard TLS handshake, as shown in Figure~\ref{fig:throughput-local-stability}. This stable impact to throughput is also observed across different levels of client concurrency.

\begin{figure*}[t]
  \centering
  \begin{subfigure}[b]{0.33\textwidth}
    \includegraphics[width=0.99\columnwidth]{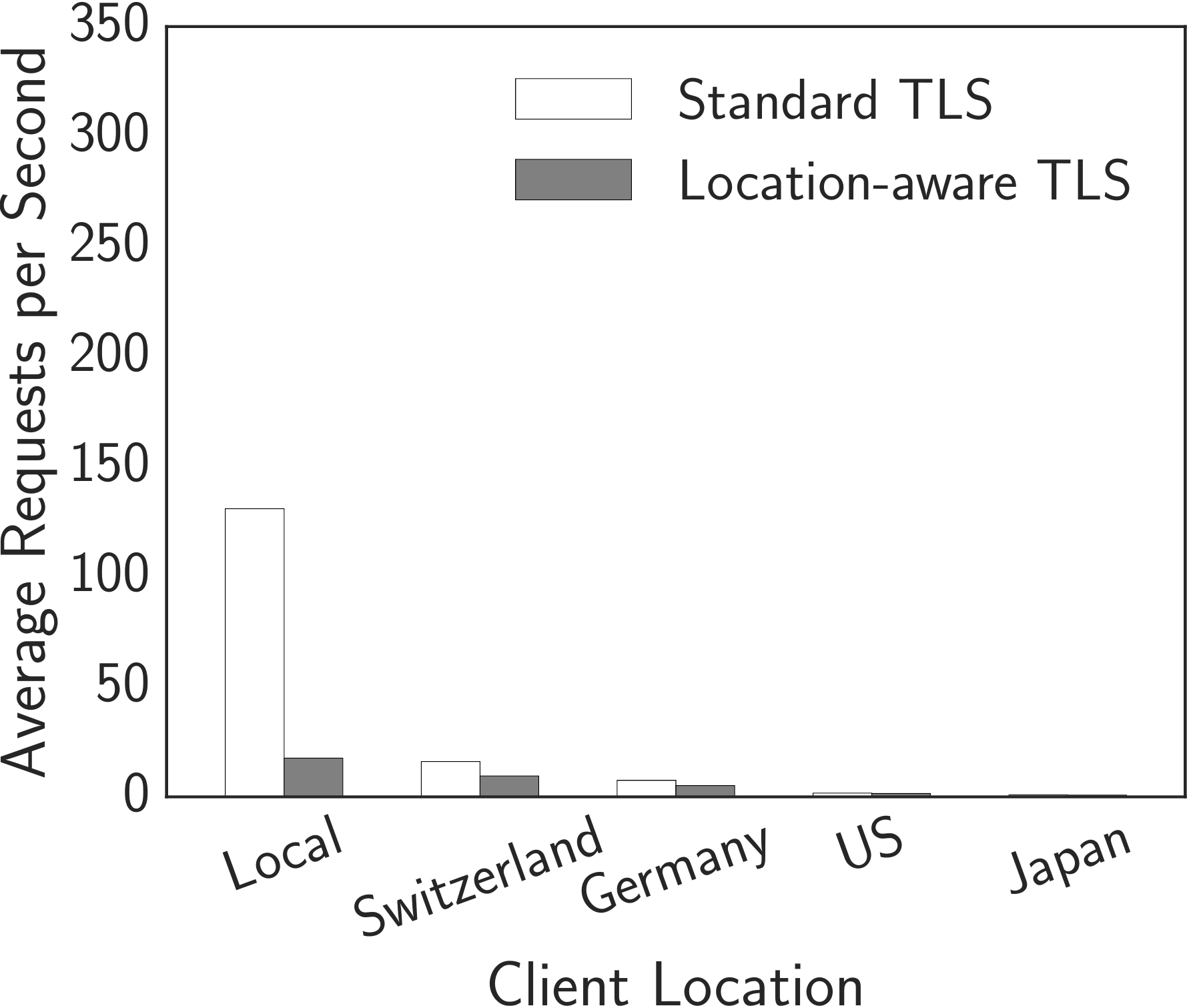}
    \caption{1 concurrent client}
  \end{subfigure}%
  \begin{subfigure}[b]{0.33\textwidth}
    \includegraphics[width=0.99\columnwidth]{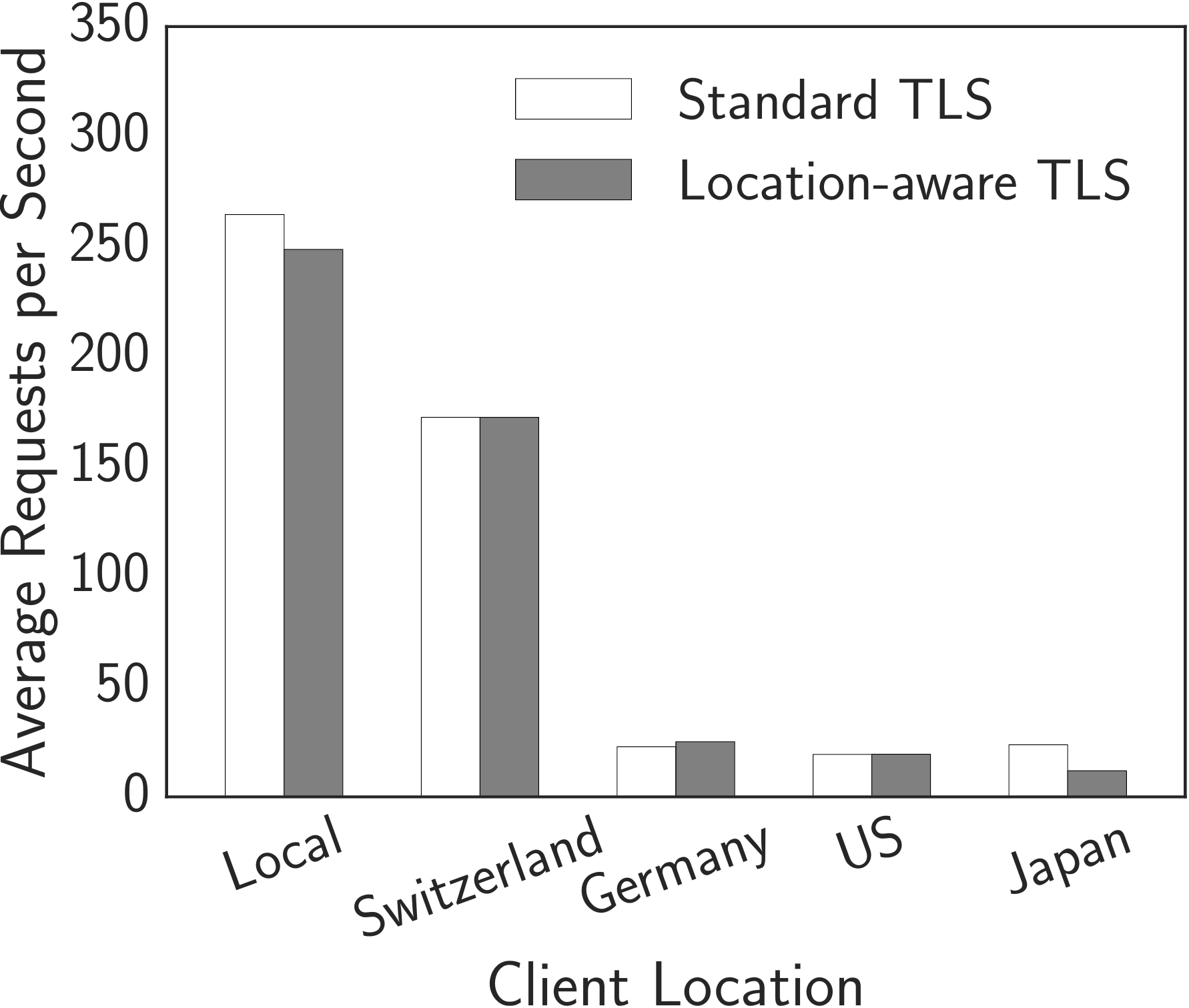}
    \caption{100 concurrent clients}
  \end{subfigure}%
  \begin{subfigure}[b]{0.33\textwidth}
    \includegraphics[width=0.99\columnwidth]{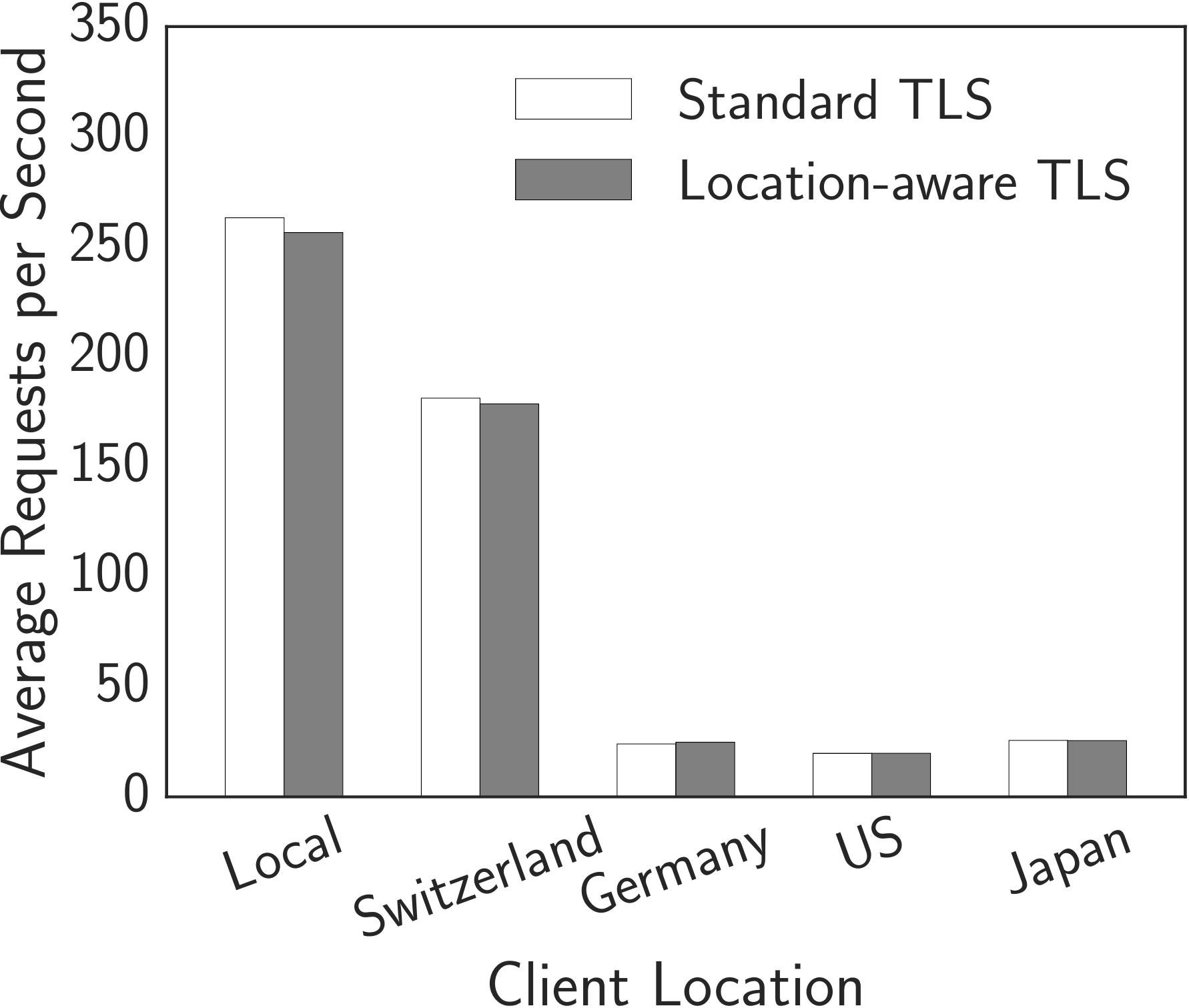}
    \caption{200 concurrent clients}
  \end{subfigure}%
  \\[6pt]
  \begin{subfigure}[b]{0.33\textwidth}
    \includegraphics[width=0.99\columnwidth]{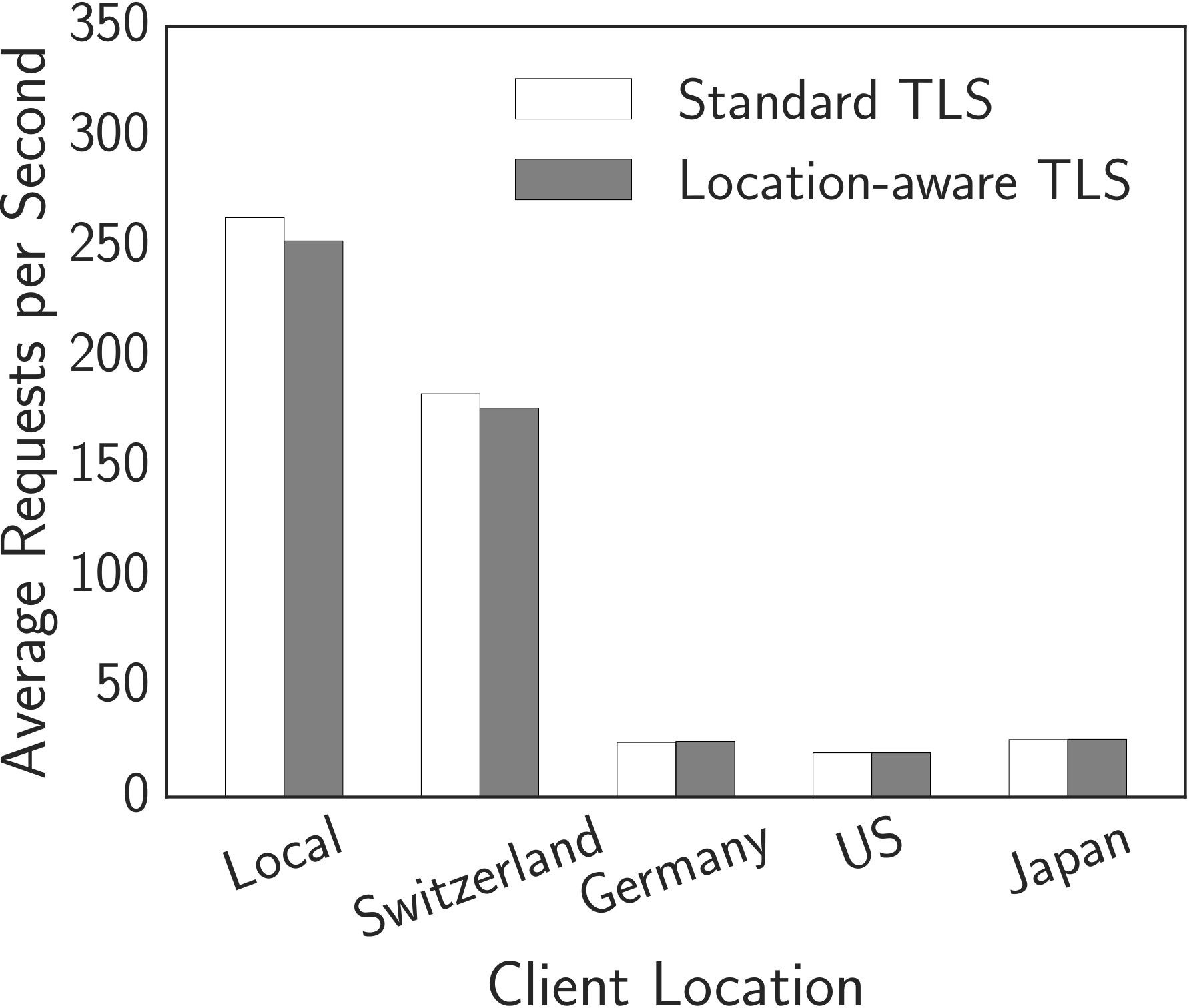}
    \caption{300 concurrent clients}
  \end{subfigure}%
  \begin{subfigure}[b]{0.33\textwidth}
    \includegraphics[width=0.99\columnwidth]{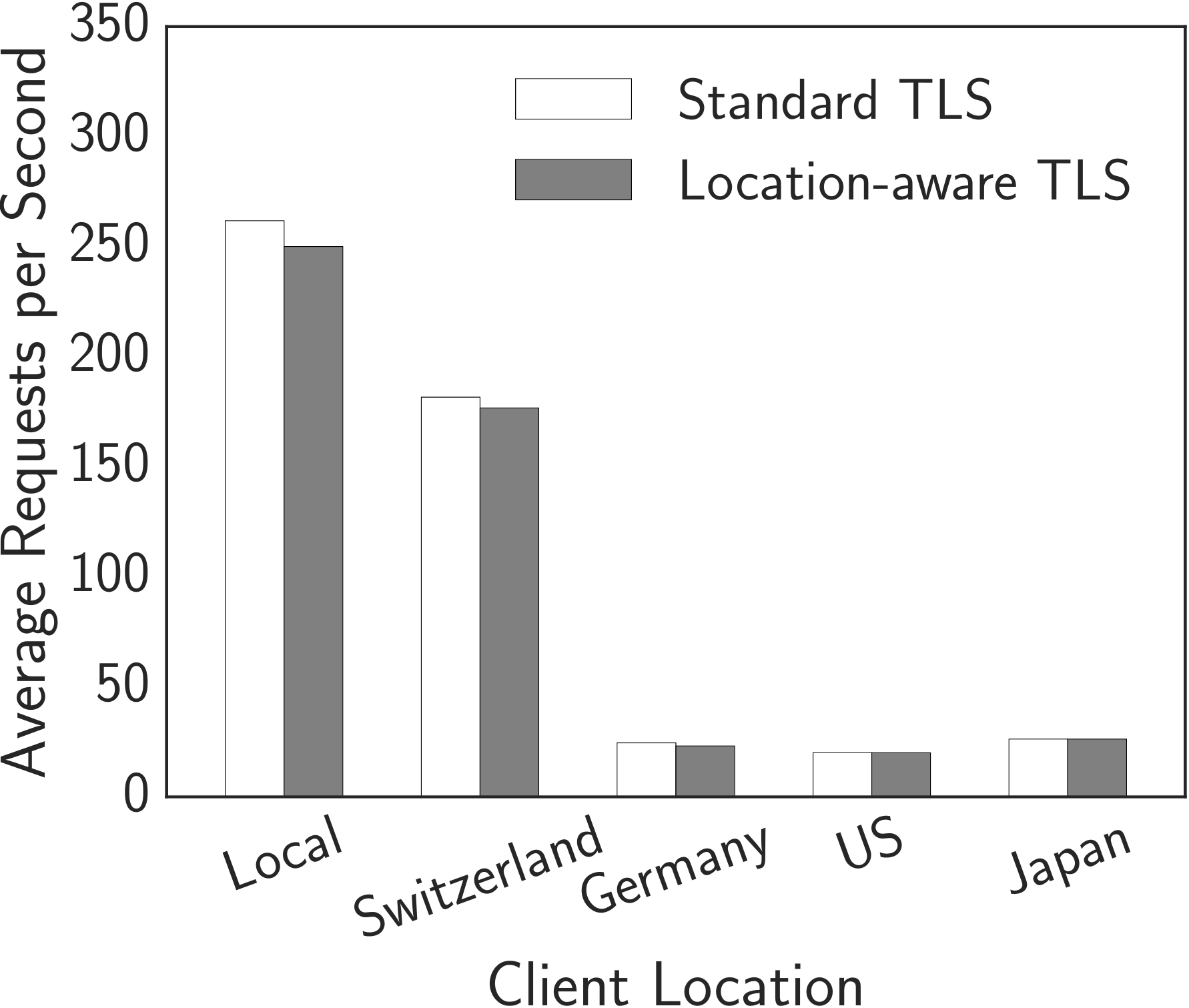}
    \caption{400 concurrent clients}
  \end{subfigure}%
  \begin{subfigure}[b]{0.33\textwidth}
    \includegraphics[width=0.99\columnwidth]{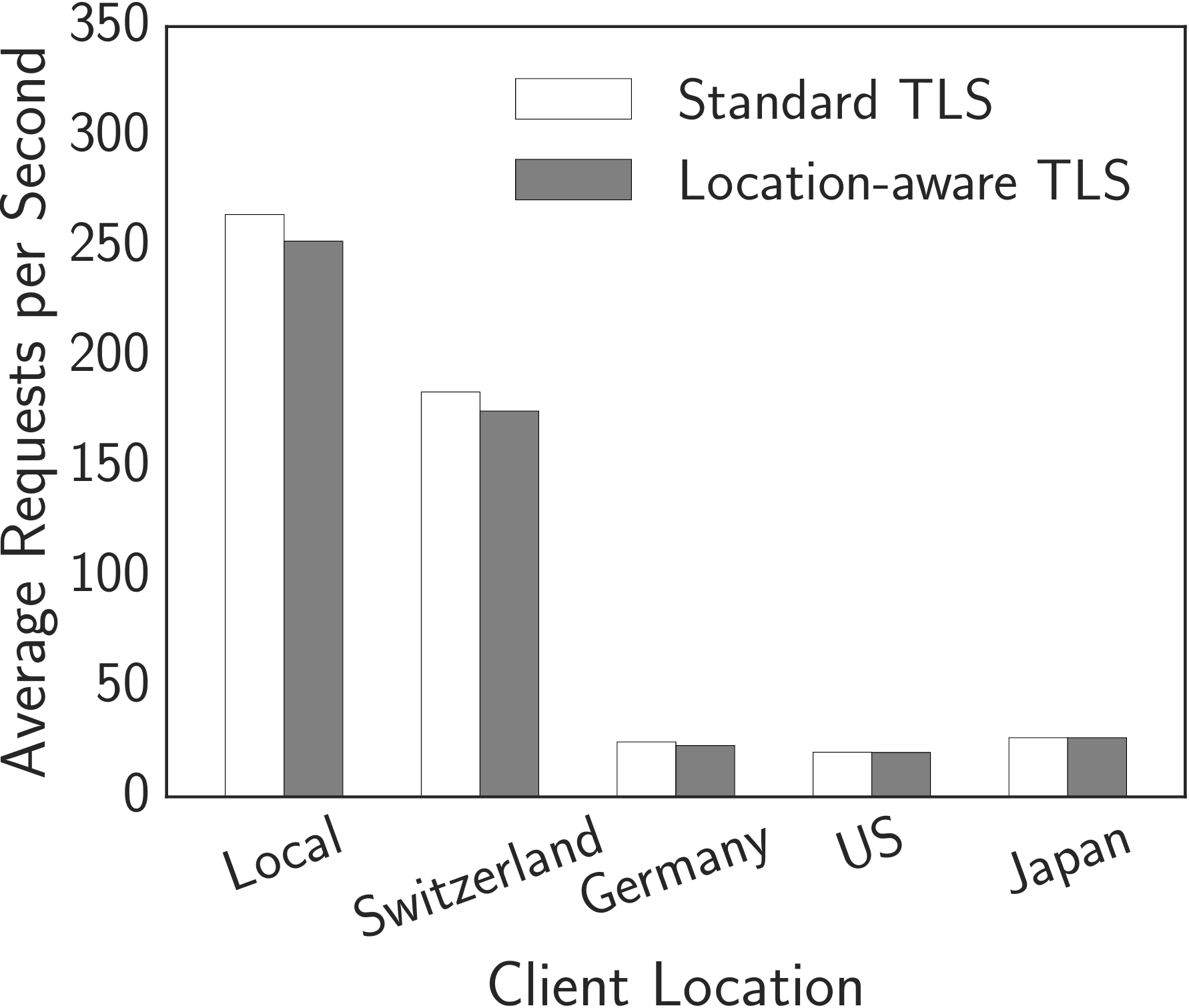}
    \caption{500 concurrent clients}
  \end{subfigure}%
  \caption{Server-side TLS handshake throughput for different number of concurrent clients evaluated from different nodes using PlanetLab. For all client concurrency levels, the difference between the throughput of \latls{} and standard TLS handshakes becomes unnoticeable as the client is geographically more distant from the server.}
  \label{fig:planet-throughput}
\end{figure*}

Since server throughput is a key indicator for website performance, we further evaluate \archname{} on a global scale using PlanetLab. We observe that the server does not experience a noticeable performance penalty for remote clients. As shown in Figure~\ref{fig:planet-throughput}, the server handles connections from Germany, the US, and Japan with little or unnoticeable impact on throughput. This is due to the network delays between the client and the server, which overshadow the overhead due to the server-to-GMLC communication.

\inlinetitle{Client-side Verification.}
The verification of the location statement done on the client side is efficient. We measure the time required for the client to validate the location statement in our modified Chromium browser. This includes checking the GMLC's signature on the location statement, verifying that the hash of the master secret matches the one computed locally on the master secret of the channel, and determining whether the location matches the legitimate server location fetched via \ladns{}. It takes, on average, 132 microseconds (standard deviation 40 microseconds) to verify a location statement. Since web page loading times are typically on the order of a few seconds~\cite{globalsitespeed}, the sub-millisecond overhead of this verification is negligible for most web applications.

\section{Discussion}
\label{sec:discussions}

We now discuss the limitations of \archname{} and possible improvements. We then explore deployment and the integration with other web mechanisms and protocols.

\subsection{Limitations}
\label{sec:limitations}

\archname{} has some limitations due to the use of various components in our system, from server deployment to the current state of LCS.

\inlinetitle{IaaS Clouds.}
This work is partly motivated by the trend of using private data centers for security-critical applications.
However, a web service may also be deployed in IaaS clouds like Amazon EC2.
IaaS cloud deployments make it easier for an attacker to be co-located with a target web service.
If the IaaS provider allows any customers to deploy SIM cards in its data centers, an adversary may appear at a legitimate location of a target web service.
Therefore, \archname{} is only suited for security-critical applications hosted at private sites that are difficult for an adversary to infiltrate.

Co-location can be prevented as a side-benefit of using reverse proxies such as NGINX~\cite{moreno2009elastic, nginx}, which are used to offload TLS encryption from web servers in the public cloud. Since reverse proxies are often deployed in private data centers with on-site security, the attacker cannot co-locate with them and our solution applies.

\inlinetitle{Accuracy and precision of Location Estimation.}
The location of the server's SIM is estimated by LCS using techniques such as Cell ID, U-TDOA, or Enhanced Cell IDs. The accuracy and precision of these localization techniques directly affects the security of \archname{}.
If localization is coarse-grained, an attacker may easily be located at a position that would be considered valid by the client.
This is a risk for servers located in remote or suburban areas where location estimation may be limited due to fewer surrounding base stations. However, telecommunication operators often provide femtocells to increase cellular coverage. Such hardware can therefore improve localization accuracy and precision.
In any case, our proposal benefits from the MLP's abstraction of the underlying localization method, allowing future improved techniques to be used when they become available.

\inlinetitle{Security of Localization Methods.}
Throughout the paper, we assume the legitimate web server (and its SIM) to be trusted. In the case of a dishonest server that tries to fake its location, secure localization methods based on distance bounding~\cite{brands1993distance,capkun2005secure} can be used to provide verifiable locations. Device locations can also be estimated using base stations with undisclosed locations~\cite{Capkun2006}.

%

\subsection{Extensions and Optimizations}
\archname{} can be further improved and adapted to fit into a larger
application space which we now describe.

\inlinetitle{Thwarting Co-Located Attackers.}
As discussed, web services are often deployed across multiple data centers. We now consider an attacker that is co-located with \emph{some} data centers, potentially due to on-site physical breach~\cite{dronedatacenter} or the inaccuracy of localization. To address this, the location statement can be extended to contain \emph{all} current server locations the server's credential is allowed to request from the GMLC. More specifically, the domain owner can specify all the server's credentials to be authorized to query for all the domain's server locations from the GMLC. Upon request by the client, the server requests the GMLC for a location statement that includes the locations of \emph{all} the SIMs of all the servers where the web service is deployed.
The GMLC checks that the server is authorized to request all SIM locations in the query.
Upon receiving the location statement, the client checks that the locations of all the servers included in the statement match all the legitimate locations fetched from \ladns{}.
While this extension may incur higher verification times, the partially co-located attacker cannot impersonate the server.

\inlinetitle{Location Privacy.}
Server location may be a valuable asset which an online service provider does
not want to disclose. In scenarios where location privacy is desired, the client
should be able to check that the server is at the \emph{correct} location
without learning its exact position. A simple way to achieve location privacy is
to include coarse-grained location information in the \texttt{LOC} records.
Similarly, the GMLC should use the same coordinate resolution in the location
statement. For example, the GMLC may allow the server to specify the location
accuracy and precision with which the object of the query should be localized.
Alternatively, our solution can use anonymous
credentials~\cite{camenisch2002signature} to realize the location
statement in a privacy-preserving way.

\begin{figure}[t]
  \centering
  \includegraphics[width=\columnwidth]
  {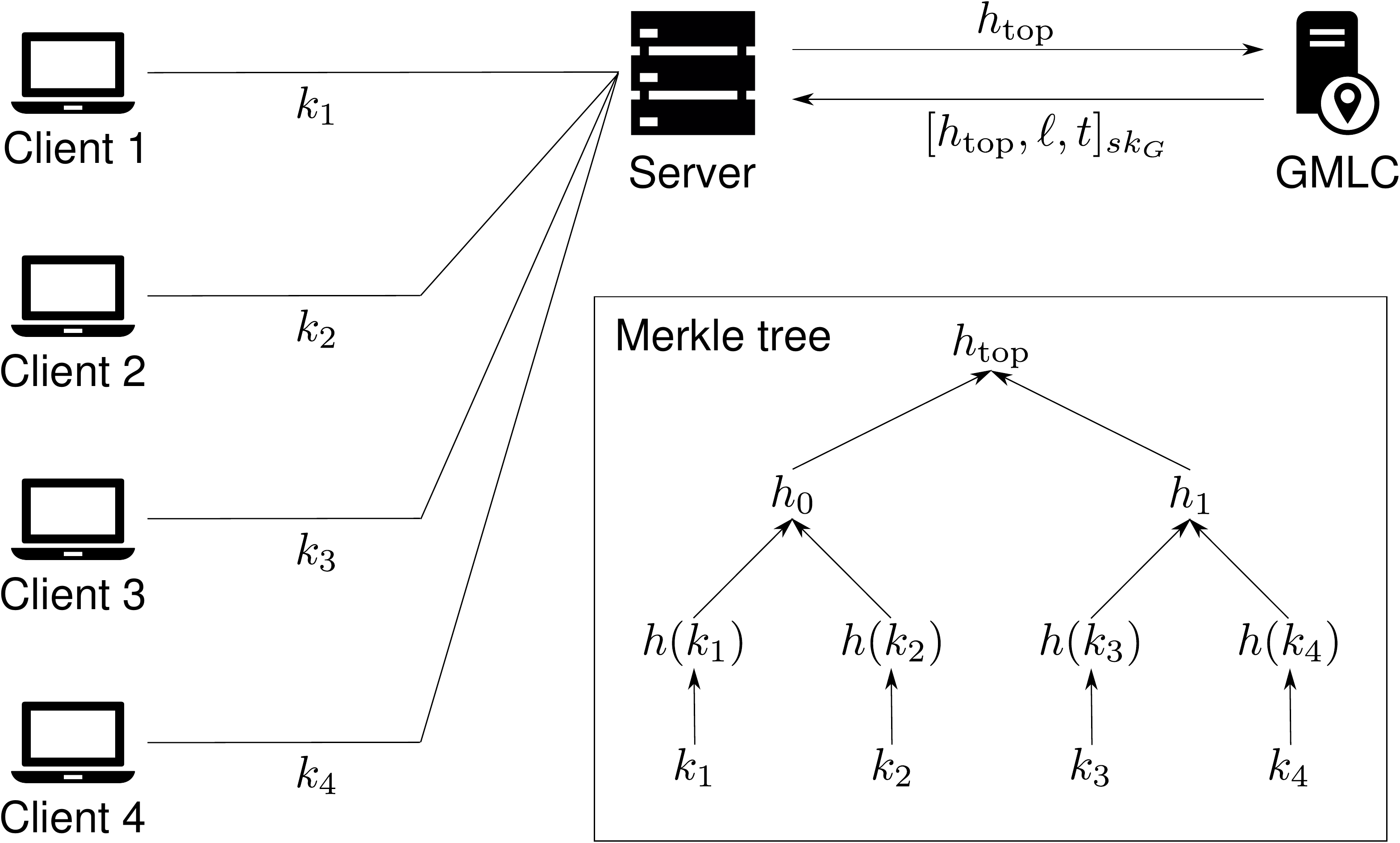}
  \caption{Improved \latls{} using aggregated location
statements. Instead of requesting the GMLC for a location statement for each
client, the server uses a Merkle tree to aggregate a set of master secrets with
the clients and requests only one statement that contains the top hash of the
tree.}
  \label{fig:location-tls-handshake-tree}
\end{figure}

\inlinetitle{Improved Concurrency Handling.}
In the current solution, the server requests a location statement for \emph{each}
individual TLS connection. While our implementation demonstrates the efficiency
of communicating with the GMLC, such a mechanism can be improved. One option is
to leverage Merkle trees~\cite{merkle1988digital} and request a single
aggregated location statement for a series of client connections. In particular,
given $N$ incoming clients in a given time frame, the server constructs a Merkle
tree where the leaf nodes are the master secret hashes of the $N$ TLS sessions.
The root hash of the Merkle tree is computed and sent to the GMLC. The resulting
location statement consists of the root hash and the server's location.
Using the \texttt{LocationStatement} message, the server sends the location
statement to each client along with the necessary sibling hash values needed by
each client to compute the root hash. As shown in
Figure~\ref{fig:location-tls-handshake-tree}, for example, the server sends
to Client~1 the location statement, along with the hash values $h(k_2)$ and
$h_1$, for Client~1 to compute and verify the root hash.

\inlinetitle{Prevention against Protocol Downgrades.}
Recall that during \latls{}, the server echos the \archname{} extension type in the \texttt{ServerHello} to indicate its support.
To circumvent \archname{}, the attacker who impersonates the server can leave out this extension type to claim that it does not support \latls{}. To prevent such an attack, an additional DNS record may be defined and populated on the DNS server to indicate to the client (during \ladns{}) the requirement and capability of \archname{} on the server side. Additionally, for known critical web services, the client may be configured to strictly require \latls{} and disconnect if the server does not support it.


\subsection{Deployment and Integration}

We discuss how \archname{} can be practically deployed and how it can be integrated with existing techniques.

\inlinetitle{Incremental Deployment.}
\archname{} can be incrementally deployed since all modifications are backward compatible (modifications to TLS and MLP are both implemented as extensions).
First, the Location Service upgrades its GMLC to offer location statement issuing for subscribed SIMs. Web services then equip their servers with SIMs and subscribe to the GMLC. They then register their server locations into DNS and install TLS libraries that support \latls{}. Finally, browser developers can implement support for \ladns{} and \latls{} to request location-based server authentication.

\archname{} is backward compatible as unsupported browsers would not request the \archname{} TLS extension. Unsupported websites would not store location information in DNS and would not respond to the \archname{} TLS extension type.
\newline


Websites use various protocol enhancements to improve the performance of TLS traffic. We discuss how \archname{} can be seamlessly integrated with them.
\newline

\inlinetitle{TLS Renegotiation.}
TLS renegotiation allows the client or the server to re-establish a TLS session
using a new TLS handshake~\cite{rescorla2009transport}. Our solution is
implemented as part of the TLS handshake and therefore supports TLS
renegotiation. If the new TLS handshake is performed over an existing TLS
session that has already verified the server's location, a
second location verification may be optional.

\inlinetitle{TLS Resume.}
\archname{} supports TLS resumption as long as the server locations do not change. Past TLS sessions can be resumed by the client, specifying the past session identifier in the \texttt{ClientHello} message~\cite{rfc5246} to the server. The session can be resumed with the same master secret if the server has the session in its cache. Upon resuming the session, the client may optionally request a new verification of the server's location by renegotiation.

\inlinetitle{TLS False Start.}
TLS False Start~\cite{langley2010transport} allows TLS endpoints to send data before the handshake completes, provided that the key exchange is TLS\_DHE or TLS\_ECDHE and the symmetric cipher used for application data is strong enough. Essentially, the \texttt{Finished} messages retroactively validates the success of the handshake. In our \latls{}, the \texttt{LocationStatement} message also acts as another retroactive validation of the handshake using location-based authentication. Therefore, the client should not send sensitive data to or authenticate data from the server before receiving and verifying the \texttt{LocationStatement}. The same rationale applies to TLS 1.3~\cite{tls13}, which allows the client to send data along with the \texttt{ClientHello} message in the 0-RTT mode, before the server's certificate is validated.

\inlinetitle{TLS Proxy.}
TLS proxies are often used to allow a network middlebox to analyze the traffic (e.g.,
for malware detection) and apply security policies~\cite{tlsproxy}. Middleboxes are
often installed on the client's gateway to the Internet and can be perceived as
a sanctioned man-in-the-middle. During the TLS handshake, the client establishes
a TLS session with the proxy, which in turn establishes a TLS session with the
server. Since it is the TLS proxy that actually connects to websites, it should be in charge of verifying the locations of all the connected servers using our approach.

\inlinetitle{Reverse Proxies.}
Some data centers often use reverse proxies at the network boundary of their
actual servers to improve efficiency~\cite{dietz2012origin}. Essentially, the
TLS session ends here and application data travels within the data center
stripped of the TLS layer. \latls{} should therefore verify the locations of the reverse proxies.
Similarly, the DNS \texttt{LOC} records should specify the locations of the
reverse proxies, rather than the ones of the actual application servers. Finally,
the SIM of the server should be installed at the reverse proxy so that the GMLC
can provide location statements that match the legitimate location distributed
via \ladns{}.

\inlinetitle{Data Center Load Balancing.}
Websites often perform load balancing across different servers within the same data center. If TLS connections terminate at these servers (instead of a TLS proxy), then all server locations should be stored in the DNS for \latls{} to succeed. As an optimization, servers in the same data center can have their SIMs installed at the same location to reduce the number of corresponding \texttt{LOC} records in DNS.

\inlinetitle{Datagram TLS.}
Datagram TLS realizes TLS on datagram traffic in UDP~\cite{rescorla2012datagram}. Since it uses the same handshake protocol, \latls{} is directly applicable to it.

\inlinetitle{SPDY and HTTP/2.}
SPDY~\cite{belshe2012spdy} and HTTP/2~\cite{http2} are new protocols that improve network efficiency by multiplexing HTTP requests to the same domain over a single shared TLS channel. SPDY also provides \emph{IP pooling}, allowing multiple HTTP sessions to use the same TLS connection for servers with the same IP address. Since \latls{} checks the location statement against the locations of the domain, server authentication succeeds when the IP address maps to the same set of legitimate server locations.

\inlinetitle{Quick UDP Internet Connections (QUIC).}
QUIC is an experimental protocol proposed by Google that achieves higher
performance over UDP~\cite{quic}. To boost efficiency, clients can
start sending data to the server, encrypted under the server's public key,
before establishing the ephemeral session key. \latls{} can be
coupled with QUIC. However, since we consider an attacker who compromises the
server's secret key, no sensitive data should be sent before the ephemeral
session key is established.

\inlinetitle{Keying Material Exporters.}
Key Material Exporters~\cite{rescorla2010keying} is a mechanism that allows secrets derived during a TLS handshake to be exported and used in the upper layer applications.
In this work, we choose to implement \archname{} within the TLS handshake since it is where the server is originally authenticated (using its certificate). 
This also relieves developers of the burden to implement location verification for each desired application.
However, we can also use key material exporters to expose secrets to the upper level applications to perform location-based server authentication.
For example, the hash of the master secret, $h(k)$, can be exposed to the server (Apache) and the client (Chromium). The server can contact the GMLC to obtain the location statement and send it over the established TLS channel to the client to be verified.

\section{Related Work}
There has been considerable research efforts on location-based
services~\cite{saroiu2009enabling,Lenders2008Location,luo2010veriplace}. They
mostly focus on proving the location on the client side, while this work uses
location to strengthen server authentication. Other
solutions exist for enhancing server authentication in TLS and we summarize them
below. We refer the reader to the work by Clark and van
Oorschot~\cite{clark2013sok} for a more detailed treatment of current HTTPS and TLS
enhancements.

\inlinetitle{Enforcing HTTPS.} To prevent attacks that steal user
credentials by using HTTP (instead of HTTPS), one could use the HTTP Strict
Transport Security. This enables the server to inform the client that only HTTPS
sessions are allowed~\cite{hodges2012http}. Furthermore, Upgrade Insecure Requests
allow websites to easily migrate to HTTPS by notifying the
client browser using the HTTP header~\cite{upgradeinsecure}.

\inlinetitle{Certificate Revocation.}
Revocation mechanisms have been proposed for wrongly issued, fraudulent, or
compromised certificates~\cite{myers1998revocatoin}. The Online Certificate
Status Protocol (OCSP) allows the client to check the validity of the
server's certificate during the TLS handshake~\cite{rfc6960}.
OCSP stapling is also proposed as a TLS extension that allows the server to send
certificate status to the client during a TLS handshake~\cite{rfc6066}.
Short-lived certificates enable the use of certificates with a shorter validity
and therefore require CAs to issue fresh certificates more
frequently~\cite{topalovic2012towards}.

\inlinetitle{Multi-path Probing.}
Multi-path probing refers to verifying a server's certificate based on evidence observed from different
independent sources. Existing examples of such
solutions include the Certificate Transparency project~\cite{rfc6962} and the
use of Perspectives~\cite{wendlandt2008perspectives}. Perspectives allows a
client to fetch records related to the server's key from trusted notaries
that store the history of keys used by that server.
The Certificate Transparency project proposes using a public log of certificates
for auditing.

\inlinetitle{Pinning.}
Key pinning solutions restrict the set of server public keys that are trusted by
a client. This prevents an attacker who obtains a fraudulent certificate over
his own public-private key pair from impersonating the server. CertLock is a
Firefox add-on that caches the information of CAs used in a
certificate~\cite{soghoian2012certified}. Web servers may also provide pinning
information to clients, as proposed in the Public Key Pinning Extension for
HTTP~\cite{rfc7469}. DANE uses DNS and DNSSEC to store the pinning of CAs, server public keys, and server
certificates~\cite{hoffman2012dns}. In general, browser developers may also
include their own key pinning information; this is implemented, for instance, in
modern browsers, such as Firefox~\cite{firefoxpinning} and
Chrome~\cite{clark2013sok}.

\inlinetitle{Comparison.}
All the above solutions address server authentication by restricting the impact
of fraudulent certificates or reducing the window of attack for the adversary.
In contrast to certificate-based approaches, we use a physical property of the server, its geographic location, to realize second-factor server authentication.
\newline

Using \ladns{} and \latls{}, \archname{} binds the server's location to the session. Such a binding is similar to the concept of TLS session-aware (TLS-SA) user authentication~\cite{oppliger2006ssl, oppliger2008ssl}. TLS-SA achieves user authentication by binding the user's credentials to the session of the authentic client; this allows the server to determine whether its session with the client is the same as that observed by the client. The difference here is that we apply such a binding on the server end and use an external party to certify the server's property (geographic location). As a result, the client can check whether the TLS session is the same as the session over which the server sends the location statement.

\section{Conclusion}
\label{sec:conclusion}

Remote web server authentication has been a challenging problem. As a solution, we leveraged the server location as a second factor of its authenticity. We implemented location-based server authentication with only minor modifications to existing software. \archname{} prevents server impersonation even if a remote attacker possesses the secret key of the victim server. Our approach requires no user intervention and incurs minimal impact to real-world performance. This work motivates the future use of various properties of protected data centers for stronger server authentication guarantees.


\section*{Acknowledgment}
We thank David Basin, Marco Guarnieri, Thomas Locher, and Hossein Shafagh for their insightful feedback and suggestions.
We also thank the anonymous shepherd and reviewers for their helpful comments to improve this work. 
Der-Yeuan Yu was funded by ABB Corporate Research.
Aanjhan Ranganathan was supported by the Zurich Information Security and Privacy Center.
Claudio Soriente was partially funded by the European Commission in part of the ReCRED project
(Horizon H2020 Framework Programme of the European Union under GA number 653417).
The paper represents the views of the authors. 
Some figure icons were made by Freepik and Picol from www.flaticon.com.

\bibliographystyle{abbrv}
\bibliography{salve}

\end{document}